\documentclass[prd,twocolumn,showpacs]{revtex4}
\usepackage{epsfig}
\usepackage{graphicx}
\usepackage{amsmath}
\usepackage{color}
\usepackage{longtable}
\usepackage[OT2,T1]{fontenc}

\newcommand{\citepaperi}{astro-ph/0702143}
\newcommand{\citepaperiii}{astro-ph/0702145}
\newcommand{\paperiatomic}{A}
\newcommand{\appprob}{D}

\newcommand{\sha}{{\text{\fontencoding{OT2}\selectfont SH}}}

\newcommand{\nucut}{\Delta\nu_{\rm cut}}
\newcommand{\rmax}{{r_{\rm max}}}

\newcommand{\gammabb}{\gamma_{\rm bb}}
\newcommand{\gammadist}{\gamma_{\rm dist}}
\newcommand{\gammaspon}{\gamma_{\rm spon}}
\newcommand{\gammastim}{\gamma_{\rm stim}}

\newcommand{\fcyc}{f_{\rm cyc}}
\newcommand{\kB}{k_{\rm B}}

\newcommand{\Tr}{T_{\rm r}}

\newcommand{\nH}{n_{\rm H}}

\newcommand{\tauS}{\tau_{\rm S}}
\newcommand{\phaL}{\pha_{\rm L}}

\newcommand{\Pesc}{P_{\rm esc}}

\newcommand{\HI}{H{\sc ~i}}

\newcommand{\HeI}{He{\sc ~i}}
\newcommand{\HeII}{He{\sc ~ii}}

\newcommand{\beq}{\begin{equation}}
\newcommand{\eeq}{\end{equation}}
\newcommand{\beqa}{\begin{eqnarray}}
\newcommand{\eeqa}{\end{eqnarray}}

\newcommand{\pha}{\mathcal{N}}
\newcommand{\barr}{\begin{array}}
\newcommand{\earr}{\end{array}}

\begin{document}
\title{Primordial helium recombination II: two-photon processes}
\author{Christopher M. Hirata}
\email{chirata@sns.ias.edu}
\affiliation{School of Natural Sciences, Institute for Advanced Study, Princeton, New Jersey 08540, USA}
\author{Eric R. Switzer}
\email{switzer@princeton.edu}
\affiliation{Department of Physics, Princeton University, Princeton New Jersey, 08544, USA}
\date{\today}

\begin{abstract}
Interpretation of precision measurements of the cosmic microwave background (CMB) will require a detailed understanding of the recombination era, which 
determines such quantities as the acoustic oscillation scale and the Silk damping scale.  This paper is the second in a series devoted to the subject 
of helium recombination, with a focus on two-photon processes in \HeI.  The standard treatment of these processes includes only the spontaneous 
two-photon decay from the $2^1S$ level.  We extend this treatment by including five additional effects, some of which have been suggested in recent 
papers but whose impact on \HeI\ recombination has not been fully quantified.  These are: (i) stimulated two-photon decays; (ii) two-photon absorption 
of redshifted \HeI\ line radiation; (iii) two-photon decays from highly excited levels in \HeI\ ($n^1S$ and $n^1D$, with $n\ge 3$); (iv) Raman 
scattering; and (v) the finite width of the $2^1P^o$ resonance.  We find that effect (iii) is highly suppressed when one takes into account destructive 
interference between different intermediate states contributing to the two-photon decay amplitude.  Overall, these effects are found to be 
insignificant: they modify the recombination history at the level of several parts in $10^4$.
\end{abstract}

\pacs{98.70.Vc, 95.30.Jx}
\maketitle

\section{Introduction}

The anisotropy of the cosmic microwave background (CMB) has proven to be one of the most versatile and robust cosmological probes.  The Wilkinson 
Microwave Anisotropy Probe (WMAP) satellite has recently measured these anisotropies at the percent level on degree scales \cite{2006astro.ph..3451H, 
2006astro.ph..3450P}, and several experiments are ongoing or planned to make precise measurements of the polarization and the sub-degree temperature 
fluctuations \cite{2001PhRvD..63d2001L, 2001PhRvL..86.3475J, 2002ApJ...568...46P, 2002MNRAS.334...11A, 2004AdSpR..34..491T, 2003NewAR..47..939K, 
2004SPIE.5498...11R,2004ApJ...600...32K, 2001ApJ...561L...7S, 2002ApJ...571..604N, 2002ApJ...568...38H, 2003ApJ...591..556P, 2002MNRAS.329..890G, 
2005MNRAS.363...79G}.  The CMB data, whose precision and robustness are so far unmatched by low-redshift observations, have provided some of the 
strongest tests of the standard cosmological model, including the adiabaticity and Gaussianity of the primordial perturbations and the spatial flatness 
of the universe.  Most recently, the CMB has provided intriguing evidence for departure of the spectrum of the primordial perturbations from scale 
invariance ($n_s<1$), as predicted by many models of inflation \cite{2006astro.ph..3449S}.

The robustness of the CMB stems from the fact that the primary anisotropy can be calculated from first principles with reasonable computing time to a 
numerical accuracy of $\sim 0.1\%$ (i.e. good enough that this is not a limiting factor) \cite{2003PhRvD..68h3507S}.  The major exception to this 
statement is recombination, which affects the CMB anisotropy because it determines the Thomson opacity and the visibility function.  The subject of 
cosmological recombination has a long history, with the early simple approximations \cite{1968ApJ...153....1P, 1968ZhETF..55..278Z} being replaced by 
more sophistocated radiative transfer and multi-level atom analyses \cite{1989ApJ...338..594K, 1990ApJ...353...21K, 1994ApJ...427..603R, 
1999ApJ...523L...1S, 2000ApJS..128..407S}.  Several papers have appeared recently suggesting that the treatment of recombination in the current 
generation of CMB anisotropy codes \cite{1999ApJ...523L...1S, 2000ApJS..128..407S} is incomplete \cite{2006A&A...446...39C, 2005AstL...31..359D, 
2004MNRAS.349..632L, 2006astro.ph.10691W, 2006AstL...32..795K} and that the remaining errors may be large enough to be relevant for next-generation 
experiments such as Planck \cite{2006MNRAS.373..561L}.  It is especially worrisome that some of the errors in the standard recombination history, in 
particular helium recombination, are partially degenerate with the scalar spectral index $n_s$, a key parameter for constraining models of inflation 
\cite{1997PhRvD..56.3207D, 2003ApJS..148..213P}.  It is therefore necessary to take a fresh look at the recombination problem.

This paper (``Paper II'') is the second in a series devoted to cosmological helium recombination.  The first of these is Switzer \& Hirata \citepaperi, 
hereafter ``Paper I,'' which re-examined helium recombination, taking into account the effects of semiforbidden and forbidden transitions, spectral 
distortion feedback, and \HI\ bound-free continuum opacity.  We believe these are the major effects in helium recombination that are not included in the 
standard treatment.  This paper considers several revisions to the standard treatment of two-photon transitions; these revisions do not have a major 
influence on helium recombination, but need to be included in order to establish that they are not important.  The emphasis is on helium although some 
of the discussion (particularly that in Secs.~\ref{sec:twophotontwo} and \ref{sec:3}) also applies to hydrogen.  The third paper of the series (Switzer \& Hirata 
\citepaperiii, hereafter ``Paper III'') will consider the effects of $^3$He scattering, electron scattering, rare decays, collsions, and peculiar 
velocities and summarize the major results.

The standard treatment of two-photon transitions in helium includes only the spontaneous two-photon decays from \HeI\ $2^1S$ to the ground level 
$1^1S$, and their inverse process, two-photon absorption.  The first correction considered in this paper is stimulated two-photon decay of $2^1S$ to 
$1^1S$, which was first analyzed by Chluba \& Sunyaev \cite{2006A&A...446...39C} in the context of \HI\ recombination.  We reanalyze the effect here 
and also include two-photon absorption of the spectral distortion as suggested by Kholupenko \& Ivanchik \cite{2006AstL...32..795K}, which delays 
recombination by re-exciting atoms.  The stimulated two-photon transitions and absorption of the spectral distortion are found to play no significant 
role in \HeI\ or \HeII\ recombination, producing corrections to $x_e$ of the order of a few times $10^{-5}$.

The second correction considered is the two-photon decay from highly excited levels ($n^1L$, where $n\ge3$ and $L=0,2$), which was responsible for the 
largest correction to recombination in the recent paper by Dubrovich \& Grachev \cite{2005AstL...31..359D}, hereafter DG05.  The treatment of such 
decays is a subtle issue because the two-photon spectrum from (for example) $3^1D\rightarrow 1^1S$ contains a resonance associated with the allowed 
sequence of one-photon transitions $3^1D\rightarrow 2^1P^o\rightarrow 1^1S$.  Since these ``1+1'' decays are already included in the level code one 
must be careful to distinguish which parts of the two-photon spectrum should be added into the level code and which parts should be left out to 
avoid double-counting the rate.  DG05 circumvented this difficulty by excluding the intermediate states associated with energetically allowed 1+1 
decays.  This clearly avoids the double-counting problem, but of course it affects the accuracy of the computed two-photon spectrum: in 
Sec.~\ref{sec:3} we will see that, particularly for large $n$, DG05 overestimated the two-photon rate because they neglect destructive interference 
from the various intermediate states.

In order to correctly implement two-photon rates from $n\ge 3$ in a level code, we must recall why they could be important even though they are much 
slower than the 1+1 decays.  The physical reason is that in a 1+1 decay, the higher photon is emitted in a \HeI\ ${n'}^1P^o$--$1^1S$ line, and is 
likely to immediately re-excite another atom.  There is no net production of the ground state He$(1^1S)$ except in the unlikely circumstance that the 
photon redshifts out of the line or is absorbed by \HI\ before it excites a \HeI\ atom.  In contrast, the nonresonant two-photon decays in which 
neither photon is emitted within a \HeI\ line will produce a net gain of one ground state helium atom (except for the subtlety that one of 
the photons could later redshift into a \HeI\ line).  Therefore, for the purposes of the level code, the way to distinguish ``resonant'' (1+1) from 
``nonresonant'' decays is {\em not} to make the distinction based on which intermediate state appears in the decay amplitude, but rather to impose a 
cutoff in frequency space: decays in which one of the photons is within $\nucut$ of a \HeI\ ${n'}^1P^o$--$1^1S$ line are treated as resonant (1+1), and 
the rest are nonresonant.  The choice of $\nucut$ (described in Sec.~\ref{sec:nonres}) is arbitrary, reflecting the fact that the 1+1 decay is not a 
distinct physical process from two-photon decay -- rather, the damping tails of the \HeI\ ${n'}^1P^o$--$1^1S$ line merge smoothly with the two-photon 
continua from all initial states that can decay to ${n'}^1P^o$.

Our approach to considering two-photon decays in this paper is to first consider the nonresonant decays for our choice of $\nucut$, and set an upper 
bound on how much they can speed up \HeI\ recombination by neglecting re-absorption of the nonresonant photons.  The resonant two-photon decays (and 
the related processes of resonant two-photon absorption and resonant Raman scattering) can be considered as an alteration to the line profile of \HeI\ 
${n'}^1P^o$--$1^1S$, which is no longer well-described by a Voigt profile if one goes far enough out into the damping wings.  Further, the 
$2^1P^o$--$1^1S$ line now has a significant linewidth: for our choice of $\nucut$, it requires $\sim 0.02$ Hubble times for a photon to redshift 
through the line (i.e. to redshift from frequency $\nu_{\rm line}+\nucut$ to $\nu_{\rm line}-\nucut$).  Because of this, one must be careful about 
assuming the radiation field within the line is in steady state.  All of these issues will be considered in Sec.~\ref{sec:res}.

The outline in this paper is as follows.  In Sec.~\ref{sec:twophotontwo}, we consider the effect of stimulated two-photon decays from the $n=2$ level 
($2^1S$) in \HeI\ and a related process, two-photon absorption of the spectral distortion.
In Sec.~\ref{sec:3}, we discuss the two-photon decay rates from highly excited levels in \HeI\ ($n\ge3$) and 
show that they were significantly overestimated by DG05.  In order to evaluate the importance of the two-photon rates, we separate the two-photon 
spectrum into ``nonresonant'' and ``resonant'' pieces.  The nonresonant contribution is considered in Sec.~\ref{sec:nonres}, and the resonant 
contribution in Sec.~\ref{sec:res}.  We conclude in Sec.~\ref{sec:conc}.

The notation in this paper is consistent with that in Paper I, but there are several new additions.  Here we will denote the Rydberg constant by ${\cal 
R}$.  The reduced matrix element of a spin $k$ tensor operator $\langle j'||T^{(k)}||j\rangle$ is defined in accordance with 
Ref.~\cite{1960amqm.book.....E}.  (This differs by a factor of $i^k$ from Ref.~\cite{1971rqt..book.....B}, but is more convenient for our purposes 
because it makes the matrix elements real.)  We will also use the symbol $L_>\equiv\max(L,1)$, which makes many appearances in our matrix elements.  
Spontaneous two-photon decay rates will be denoted by $\Lambda$, while the finite-temperature rates will be denoted $\Gamma_{2\gamma}$.  Differential 
rates as a function of photon frequency or energy will be written $d\Gamma_{2\gamma}/d\nu$ or $d\Gamma_{2\gamma}/dE$.

\section{Two-photon decays from $n=2$}
\label{sec:twophotontwo}

The two-photon transitions from the metastable \HI\ $2s$ and \HeI\ $2^1S$ levels are an important contribution to the recombination rates.  It is 
usually assumed that stimulated two-photon emission plays a negligible role in the decay of the $n=2$ states in \HI\ and \HeI\ 
\cite{2000ApJS..128..407S}.  Chluba \& Sunyaev \cite{2006A&A...446...39C} found that stimulated emission in \HI\ $2s\rightarrow 1s$ modifies TT and TE 
anisotropies at the percent level on small scales.  We use a similar method to include \HeI\ and \HeII\ stimulated two photon emission in addition to 
\HI.  We also include the re-absorption of the spectral distortion via two-photon excitation \cite{2006AstL...32..795K}.  We do not find any 
significant effects in \HeI\ or \HeII.

In this section, we present a general treatment of the two-photon decays from $n=2$ levels, including stimulated emission and the effects of absorption of the spectral distortion.  For \HeI\ $n=2$ states, we consider first the usual two-photon decay:
\beq
\mathrm{He}(2^1S) \rightarrow \mathrm{He}(1^1S) + \gammaspon  + \gammaspon
\label{eq:i}
\eeq
and the two-photon excitation
\beq
\mathrm{He}(1^1S) + \gammabb + \gammabb \rightarrow \mathrm{He}(2^1S),
\label{eq:iv}
\eeq
where $\gammaspon$ refers to a spontaneously emitted photon and $\gammabb$ refers to a photon drawn from the blackbody radiation.  These two equations 
are the only ones considered in standard recombination codes, and they are typically included with a rate coefficient of $\Lambda_{\rm 
HeI}=51\,$s$^{-1}$ for Eq.~(\ref{eq:i}) and the detailed balance rate $\Lambda_{\rm HeI}e^{-\Delta E/\kB\Tr}$ for Eq.~(\ref{eq:iv}), where $\Delta 
E=E(2^1S)-E(1^1S)$.  As pointed out for \HI\ by Chluba \& Sunyaev \cite{2006A&A...446...39C}, one should also consider the analogous stimulated decays 
in \HeI:
\beq
\mathrm{He}(2^1S) \rightarrow \mathrm{He}(1^1S) + \gammaspon + \gammastim
\label{eq:ii}
\eeq
and
\beq
\mathrm{He}(2^1S) \rightarrow \mathrm{He}(1^1S) + \gammastim + \gammastim,
\label{eq:iii}
\eeq
where $\gammastim$ means that the photon's emission is stimulated.  [Chluba \& Sunyaev \cite{2006A&A...446...39C} replaced $\Lambda_{\rm HI}$ in their 
level code with the sum of rates for Eqs.~(\ref{eq:i}), (\ref{eq:ii}), and (\ref{eq:iii}).]  Note that it is not self-consistent to leave out these 
reactions, since the reverse reaction of Eq.~(\ref{eq:iv}) is not just Eq.~(\ref{eq:i}) but rather the combination of Eqs.~(\ref{eq:i}), (\ref{eq:ii}), 
and (\ref{eq:iii}).

If there is a spectral distortion from redshifted \HeI\ line photons, one should also consider the possibility of two-photon absorption of a thermal 
photon and a distortion photon \cite{2006AstL...32..795K}:
\beq
\mathrm{He}(1^1S) + \gammabb + \gammadist \rightarrow \mathrm{He}(2^1S),
\label{eq:v}
\eeq
where $\gammadist$ refers to a spectral distortion photon.  In principle, there is an additional contribution where both absorbed photons come from the 
spectral distortion.  This is negligible since the blackbody spectrum dominates over the spectral distortion for photons with $h\nu<\Delta E(2^1S)/2$ 
except at $z<1400$ when \HeI\ recombination is finished ($x_{\rm HeII} < 10^{-14}$).  All of these equations 
have analogues in \HI\ and \HeII.

The two-photon decay rate is
\begin{eqnarray}
\dot x_{2^1S\rightarrow 1^1S} \!\!&=&\!\! \Lambda_{\rm HeI}\int_0^{\nu_{1^1S-2^1S}/2} \phi(\nu) 
\nonumber \\ && \times \biggl\{ x_{2^1S} [1+\pha(\nu)][1+\pha(\nu')]
\nonumber \\ && - x_{1^1S} \pha(\nu) \pha(\nu') \biggr\} \,d\nu,
\label{eq:twophorate1}
\end{eqnarray}
where $\phi(\nu)$ is the two-photon emission profile, normalized to
\beq
\int_0^{\nu_{1^1S-2^1S}/2} \phi(\nu)\,d\nu=1,
\eeq
and the frequency of the higher-frequency photon is $\nu'=\nu_{1^1S-2^1S}-\nu$.  Note that decay term $[1+\pha(\nu)][1+\pha(\nu')]$ can be expanded to 
give a spontaneous piece, a singly stimulated piece $\pha(\nu)+\pha(\nu')$, and a doubly stimulated piece $\pha(\nu)\pha(\nu')$.  The phase space 
density for the higher-energy photon is much less than unity: at $z=2600$ it is $2\times 10^{-4}$ at the midpoint of the \HI\ spectrum, 
$\nu_{1s-2s}/2$, and it is even less above the midpoint, for \HeI, or for lower redshifts).  Therefore we make the replacement in the downward rate 
$1+\pha(\nu')\rightarrow 1$, i.e. we neglect stimulated emission of the higher-energy photon.  Similarly since the spectral distortion phase space 
density is $\ll 1$, we may replace $1+\pha(\nu)\rightarrow 1+\pha_{\rm bb}(\nu)$.  This enables us to write Eq.~(\ref{eq:twophorate1}) as
\beq
\dot x_{2^1S\rightarrow 1^1S} = \dot x_{2^1S\rightarrow 1^1S}^{({\rm thermal})} + \dot x_{2^1S\rightarrow 1^1S}^{({\rm nonthermal})},
\eeq
where
\beqa
\dot x_{2^1S\rightarrow 1^1S}^{({\rm thermal})} \!\!&=&\!\! \Lambda_{\rm HeI}\int_0^{\nu_{1^1S-2^1S}/2} \frac{\phi(\nu)\,d\nu}{1-e^{-h\nu/\kB\Tr}}
\nonumber \\ && \!\!\times
\left( x_{2^1S}-x_{1^1S}e^{-h\nu_{1^1S-2^1S}/kT_r} \right)
\label{eq:thermal}
\eeqa
and
\beqa
\dot x_{2^1S\rightarrow 1^1S}^{({\rm nonthermal})} \!\!&=&\!\! -\Lambda_{\rm HeI}\int_0^{\nu_{1^1S-2^1S}/2} \frac{\phi(\nu)\,d\nu}{e^{h\nu/kT_r}-1}
\nonumber \\ && \!\!\times
\pha_{\rm dist}(\nu')x_{1^1S}.
\label{eq:nonthermal}
\eeqa
[Here $\pha_{\rm dist}(\nu')$ is the distortion phase space density defined by taking the actual phase space density and subtracting the blackbody 
contribution.]

In the level code, the profiles for \HI\ and \HeII\ are based on the fits by Nussbaumer and Schmutz \cite{1984A&A...138..495N}, and for \HeI\ we use the fit to Drake \cite{1986PhRvA..34.2871D} described in Appendix~\paperiatomic\ of Paper I.

\begin{figure}
\includegraphics[angle=-90,width=3.2in]{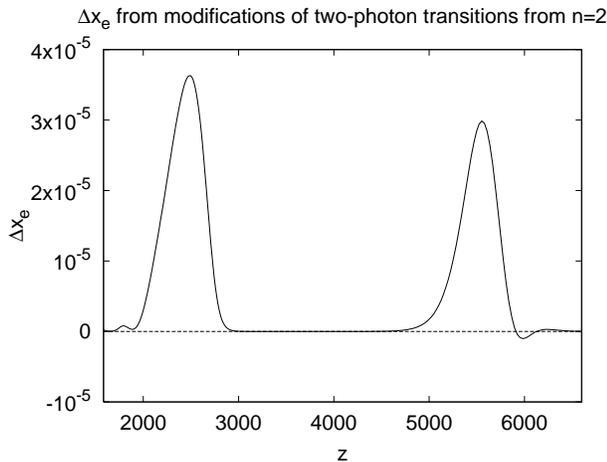}
\caption{\label{figs:twophostim}A comparison of the effect of stimulated two-photon emission and nonthermal two-photon absorption relative 
to the reference model.  The net effect is a delay in recombination, $\Delta x_e>0$.  The two peaks correspond to the effect on \HeII\ recombination 
($z\sim 5500$) and \HeI\ recombination ($z\sim 2500$).  Note that in both cases the effect on the recombination history is small, i.e. a few parts in 
$10^5$.}
\end{figure}

The results of including Eqs.~(\ref{eq:thermal}) and (\ref{eq:nonthermal}) in the level code are shown in Fig.~\ref{figs:twophostim}.  We can see that 
for \HeI\ and \HeII\ the effect is very small -- only a few times $10^{-5}$.  A larger effect in $\Delta x_e$ during hydrogen recombination was found 
by \cite{2006AstL...32..795K}, 
for several reasons.  First, the absolute abundance of hydrogen is greater, so a similar fractional change in its recombination history leads to a 
larger effect.  Second, the $2s$ and $2p$ levels in \HI\ are essentially degenerate, whereas in \HeI\ the $2^1P^o$ level lies 0.6 eV above the $2^1S$ 
level.  This changes the shape of the two-photon spectrum at low frequencies, where $\phi(\nu)\propto \nu^3$ in \HeI\ (due purely to the available 
phase space 
for emitting a low energy photon) as opposed to $\phi(\nu)\propto\nu$ in \HI\ (where there is a pole corresponding to the $2p$ intermediate state in 
the matrix element at zero frequency).  Stimulated emission and re-absorption of the spectral distortion will play a larger role in the case of \HI\ 
where the two-photon spectrum has more probability at the ends of the spectrum.  A third reason is that, due largely to the lower abundance of helium 
versus hydrogen, the \HeI\ $2^1P^o$--$1^1S$ optical depth during \HeI\ recombination is much less than that of Ly$\alpha$ during \HI\ recombination; 
therefore the importance of two-photon decays relative to resonance escape is less for \HeI\ than for \HI.

\section{Two-photon rates from $n\ge 3$}
\label{sec:3}

In this section, we consider the effect of two-photon decays from the higher excited levels of \HeI\ ($n\ge3$).  Such decays were discussed by DG05 as 
a potential means to dramatically speed up \HeI\ recombination.  The inclusion of these decays introduces a new subtlety, however, which is not present 
for the $n=2$ two-photon decay $2^1S\rightarrow 1^1S$.  The $2^1S$ level does not have any allowed decay routes, so it is correct to take a multilevel 
atom code and add in a new rate $\Lambda_{2^1S\rightarrow 1^1S}$ for this decay.  In contrast, the higher levels in \HeI\ ($n^1L$, $n\ge 3$, 
$L\in\{0,2\}$) that can undergo two-photon decay to $1^1S$ all have ``1+1'' decays in which the atom first undergoes an allowed single-photon emission 
to an intermediate level, and then undergoes a second allowed one-photon emission to reach the ground level: $n^1L\rightarrow {n'}^1P^o\rightarrow 
1^1S$ (where $2\le n'<n$).  These 1+1 decays are automatically included in the calculation of the two-photon spectrum using Fermi's Golden Rule [we 
will see this explicitly in Eq.~(\ref{eq:m})] and they turn out to dominate the net two-photon rate.  In order to include two-photon transitions from 
$n\ge 3$ levels in the multilevel atom code, we need to distinguish ``true'' two-photon decays from 1+1 decays.  It is sometimes said that in a 
two-photon decay the two photons are emitted ``simultaneously,'' but one must be careful in making this statement because the uncertainty principle 
dictates that one cannot measure the time of emission of the photons more accurately than the reciprocal of the frequency resolution.  Rather, one must 
return to the physical picture of recombination and remember that rare processes such as two-photon decay are potentially important because 
the \HeI\ resonance lines have a high optical depth and hence a high probability of re-absorption of any radiation emitted in those lines.  In contrast, 
photons emitted outside of the resonance lines have a low probability of re-absorption (unless they later redshift into a line).

Based on this picture, we can construct a ``practical'' definition for two-photon decays as follows: radiation emitted farther than some arbitrarily specified distance $\nucut$ from the nearest \HeI\ resonance line will be said to originate from a ``nonresonant two-photon decay,'' and radiation emitted within $\nucut$ of a resonance will be said to originate from a ``resonant'' or ``1+1 decay.''  The nonresonant decays exhibit a continuous spectrum and can be treated in the same way as two-photon decays from $2^1S$.  The full emission spectrum of the resonant decays is not identical to the usual Voigt profile, and the differences will have to be treated by modifying the line radiative transfer analysis.  We will consider nonresonant two-photon decays in Sec.~\ref{sec:nonres} and resonant decays in Sec.~\ref{sec:res}; the corrections to the recombination history turn out to be small in both cases.  This section will be concerned exclusively with obtaining the rate coefficients for two-photon decay, which we will find to be much less than estimated by DG05 across most of the two-photon spectrum.  This is the reason why we find only a small correction from the $n\ge 3$ two-photon decays whereas DG05 found an effect of several percent in $x_e$.

We will also consider Raman scattering from the excited levels to the ground level; the two processes, while physically distinct, are related by crossing symmetry and hence share many characteristics, including the existence of 1+1 resonances and the associated subtleties.

The outline of this section is as follows.  The formulas for two-photon decay and Raman scattering in quantum electrodynamics are introduced and summarized in Sec.~\ref{ss:2prates}.  The DG05 estimate for the rate coefficients is recalled in Sec.~\ref{ss:dg05}, and in Sec.~\ref{ss:largen} we explain why their rates are too large for the high $n$ levels.  Finally, Sec.~\ref{ss:newrate} presents our calculation of the two-photon decay rates, which are much less than those of DG05, except near resonance.

\subsection{Rates}
\label{ss:2prates}

On account of the electric dipole selection rules, two-photon transitions to the ground state of \HeI\ are allowed only from spin-singlet levels with 
even parity and $L\in\{0,1,2\}$.  In two-electron atoms only doubly excited levels can have $L=1$ and even parity, and these are inaccessible at 
recombination-era temperatures (they lie $\sim 60\,$eV above the ground state, whereas the ionization energy is 24.6$\,$eV); thus we restrict our 
attention to the $n^1L$ levels with $L\in\{0,2\}$.
Also, the two photons emerge with frequencies $\nu$ and $\nu'$ that satisfy the energy conservation condition
\beq
\nu + \nu' = \frac{E(n^1L)-E(1^1S)}h \equiv \frac{\Delta E(n^1L)}h.
\eeq
The two-photon decay rate from the $n^1L$ level to the $1^1S$ level of helium is then given by
\beq
\frac{d\Gamma}{d\nu} = \frac{\alpha^6\nu^3{\nu'}^3}{108(2L+1){\cal R}^6}
[1+\pha(\nu)][1+\pha(\nu')]|{\cal M}_{2\gamma}|^2,\;\;
\label{eq:rate-2g}
\eeq
where ${\cal R}=3.29\times 10^{15}\,$Hz is the Rydberg in frequency units, and the dimensionless amplitude is
\beqa
{\cal M}_{2\gamma} \!\! &=& \!\! a_0^{-3}
\sum_{n'} \langle 1^1S || {\bf d} || {n'}^1P^o \rangle
\langle {n'}^1P^o || {\bf d} || n^1L \rangle
\nonumber \\ && \!\! \times
\Biggl(\frac 1{\Delta E({n'}^1P^o)-h\nu}
\nonumber \\ &&
 + \frac 1{\Delta E({n'}^1P^o)-h\nu'}\Biggr),
\label{eq:m}
\eeqa
where ${\bf d}$ is the electric dipole moment operator and we have used cgs units.  [Eq.~(\ref{eq:rate-2g}) is equivalent to Eq.~(59.28) of 
Ref.~\cite{1971rqt..book.....B} after appropriate manipulation of reduced matrix elements.] Note that the summation here is over continuum levels 
with $^1P^o$ symmetry as well as discrete levels.  The total two-photon decay rate is
\beq
\Gamma_{2\gamma}(n^1L\rightarrow 1^1S) = \frac12\int_0^{\Delta E(n^1L)/h} \frac{d\Gamma}{d\nu} d\nu,
\label{eq:g2g}
\eeq
where the factor of $1/2$ occurs because we count each decay twice by integrating over the whole spectrum.

One can see that the amplitude ${\cal M}$ posesses a pole at each frequency $\nu$ corresponding to an intermediate ${n'}^1P^o$ level. 
Correspondingly, there is a branch cut (i.e. a continuous distribution of poles) for frequencies corresponding to the ${n'}^1P^o$ continuum.  Since the 
rate $\Gamma\propto |{\cal M}|^2$, the poles give rise to resonances in the cross section, which have the characteristic $\propto\nu^{-2}$ structure.  
As is usual in quantum mechanics (e.g. Sec. V\S18 of Ref.~\cite{1954qtr..book.....H}), the total rate is rendered finite by giving the energies 
$E({n'}^1P^o)$ a small imaginary part $E\rightarrow E+i\Gamma/2$, where $\Gamma$ is the width of the state.  The imaginary part changes the cross 
section in the resonance to the characteristic Lorentz form (which becomes a Voigt profile in the comoving frame due to thermal motion of the atoms).  
The resonances at $0<h\nu<\Delta E(n^1L)$ give rise to the allowed decays where the atom decays from $n^1L$ to ${n'}^1P^o$ by emission of a single 
electric dipole photon, and then proceeds to decay to $1^1S$ by emitting a second photon.  These ``1+1'' decays are in fact not distinct physical 
processes from two-photon emission.  Rather, the damping wings of the lines from 1+1 decays merge continuously into the two-photon continuum.

A phenomenon related to two-photon decay is Raman scattering from $n^1L$ to $1^1S$ through an intermediate ${n'}^1P^o$ state.  This has the same selection rules as two-photon decay.  If the incoming photon frequency is $\nu$ and the outgoing frequency is $\nu'$, we have
\beq
\nu' = \nu + \frac{\Delta E(n^1L)}h,
\eeq
and the scattering rate (in number of scatterings per atom in the $n^1L$ level per second) is
\beq
\frac{d\Gamma}{d\nu} = \frac{\alpha^6\nu^3{\nu'}^3}{108(2L+1){\cal R}^6}
\pha(\nu)[1+\pha(\nu')]|{\cal M}_{\rm Raman}|^2.
\eeq
Because of crossing symmetry, we may obtain the Raman scattering matrix element by analytic continuation of Eq.~(\ref{eq:m}) to negative frequencies:
\beqa
{\cal M}_{\rm Raman} \!\!&=&\!\! a_0^{-3}
\sum_{n'} \langle 1^1S || {\bf d} || {n'}^1P^o \rangle
\langle {n'}^1P^o || {\bf d} || n^1L \rangle
\nonumber \\ && \!\! \times
\Biggl(\frac 1{\Delta E({n'}^1P^o)+h\nu}
\nonumber \\ &&
 + \frac 1{\Delta E({n'}^1P^o)-h\nu'}\Biggr).
\label{eq:mr}
\eeqa
The total Raman scattering rate is
\beq
\Gamma_{\rm Raman} = \int_0^\infty \frac{d\Gamma}{d\nu} d\nu.
\label{eq:graman}
\eeq

The summations in Eq.~(\ref{eq:m}) and Eq.~(\ref{eq:mr}) are in general nontrivial as they depend on the helium wave functions.  Accurate calculations are available only for 
the $2^1S$ level, which is the only singlet level for which two-photon transitions are the dominant mode of decay.  In the case of cosmic recombination 
however, the ``blocking'' of allowed one-photon electric dipole decays by high line optical depth means that subdominant decay modes of highly 
excited states can become significant, and estimates of their rates are required.  DG05 was the first paper to consider these two-photon decays in the 
context of the cosmic recombination, and they introduced a simple scaling argument for the rates.  We revisit the issue here and conclude that the
decay rate is significantly smaller.

\subsection{DG05 estimate}
\label{ss:dg05}

This section reviews the derivation by DG05 of the two-photon rate from highly excited states in \HeI.  We present the key points of the derivation in the notation of this paper in order to highlight the most important assumptions in their paper and how they differ from a more detailed treatment.

DG05 noted that the dipole matrix elements of the form $\langle {n'}^1P^o || {\bf d} || n^1L \rangle$ are largest for $n'=n$.  In particular, in the limit of hydrogenic wavefunctions they show that for large $n$,
\beq
\left|\langle n^1L || {\bf d} || n^1P^o \rangle\right|^2 \sim \frac9{10}\sum_{n'}
\left|\langle n^1L || {\bf d} || {n'}^1P^o \rangle\right|^2.
\label{eq:910}
\eeq
Therefore DG05 argued that far from the 1+1 resonances in the two-photon rate, the matrix element ${\cal M}_{2\gamma}$ should be dominated by the $n'=n$ term.  They also noted the near-degeneracy of the $n^1P^o$ and $n^1L$ levels.  If one treates this degeneracy as exact, and keeps only the $n'=n$ term in the sum, one can show that the vacuum decay rate is
\beqa
\frac{d\Lambda^{\rm(DG)}}{d\nu} \!\! &=& \!\! \frac{\alpha^6\nu^3{\nu'}^3
\left|\langle n^1P^o || {\bf d} || n^1L \rangle\right|^2 
\left|\langle n^1P^o || {\bf d} || 1^1S \rangle\right|^2 
}{108(2L+1)a_0^6{\cal R}^6}
\nonumber \\ && \!\! \times
\Biggl(\frac 1{\Delta E({n'}^1P^o)-h\nu}
\nonumber \\ && 
 + \frac 1{\Delta E({n'}^1P^o)-h\nu'}\Biggr)^2,
\eeqa
where the $^{\rm(DG)}$ superscript indicates that the DG05 approximation is being used.  The frequency integral is a polynomial, and hence is trivially performed.  It results in a total decay rate of
\beqa
\Lambda^{\rm(DG)} \!\!&\propto&\!\! (2L+1)^{-1}
\left( \frac{\Delta E(n^1L)}{h\cal R} \right)^5
\nonumber \\ && \!\! \times
\left|\langle n^1P^o || {\bf d} || n^1L \rangle\right|^2 
\left|\langle n^1P^o || {\bf d} || 1^1S \rangle\right|^2.
\label{eq:dg-2g}
\eeqa
Note that within the DG05 approximation all the two-photon spectra are scaled versions of each other.  DG05 thus used Eq.~(\ref{eq:dg-2g}) to re-scale the \HI\ $2s\rightarrow 1s$ decay rate of 8.2$\,$s$^{-1}$ to the highly excited levels in hydrogen and helium, i.e. they re-scaled $\Lambda$ in proportion to the squares of the dipole matrix elements and the fifth power of the energy difference.  This leads to the result (using hydrogenic values for the $\langle n^1P^o || {\bf d} || n^1L \rangle$ matrix elements)
\beqa
\sum_{L=0,2} (2L+1)\Lambda^{\rm(DG)}(n^1L)
\!\!&=&\!\! 10540 \left( \frac{n-1}{n+1} \right)^{2n}
\nonumber\\&&\!\!\times
\frac{11n^2-41}n \;{\rm s}^{-1}.
\label{eq:dg}
\eeqa

The most important result of this is the scaling for large values of $n$.  At large values of $n$, $\Delta E(n^1L)$ approaches the ionization energy $\chi_{\rm HeI}$, whereas the dipole matrix elements scale as
$\langle n^1P^o || {\bf d} || n^1L \rangle \propto n^2$ and
$\langle n^1P^o || {\bf d} || 1^1S \rangle \propto n^{-3/2}$.
This explains the large-$n$ scaling of Eq.~(\ref{eq:dg}):
\beq
\Gamma^{\rm(DG)}_{2\gamma}(n^1L)\propto n.
\label{eq:dgscale}
\eeq
This results in a very large contribution to the two-photon rate from large values of $n$.  In fact, since the occupation probability of the large-$n$ states approaches a constant as $n\rightarrow\infty$, the {\em total} 2-photon decay rate to the ground state from highly excited helium atoms diverges as $\propto\sum n\propto n^2$ in the DG05 approximation.  DG05 cut off the sum at $n\sim 40$, since for larger $n$ the ``size'' ($\sim a_0n^2$) of the excited atom is comparable to the wavelength of the photon and hence the dipole emission formula is no longer valid.  This nevertheless leads to a very large speed-up of \HeI\ recombination.

\subsection{Large $n$ behavior}
\label{ss:largen}

Unfortunately, the simple approximation of taking only the $n'=n$ term in the summation fails for large $n$.  Indeed, it has been found for the highly
excited states of hydrogen that the actual scaling of the two-photon decay rate is $d\Lambda/d\nu\propto n^{-3}$ \cite{1987PhRvA..36.2155F}.  Here we
recall the physical argument why the scaling is $n^{-3}$, and then show that this arises due to a near-exact cancellation of matrix elements for large
$n$.  The argument has been given in a rather complicated and general form in Refs.~\cite{1979JPhB...12.3229R, 1982PhRvA..25.3079Q}, however we present
a simplified version here in order to highlight the key pieces of physics required, and see that the same argument applies to helium.

Suppose we re-write the analogue of Eq.~(\ref{eq:m}) for \HI\ in the form
\beq
{\cal M}_{2\gamma} = a_0^{-3} \left[ \langle \Psi(\nu)||{\bf d}||nl\rangle
+ \langle \Psi(\nu')||{\bf d}||nl\rangle \right],
\label{eq:mmod}
\eeq
where the states $|\Psi_m(\nu)\rangle$ are defined by \cite{1998JPhB...31.3743M}
\beq
|\Psi_m(\nu)\rangle = [H-E(1s)-h\nu]^{-1}d_m|1s\rangle.
\eeq
(Since the dipole operator $d_m$ is spin 1 it has three components $m=-1,0,+1$ and hence $|\Psi_m(\nu)\rangle$ actually consists of three states.)  Now the wave function of the $|1s\rangle$ state is localized near the nucleus, with an exponential falloff in the classically forbidden region.  The dipole operator $d_m$ simply multiplies this wave function by a polynomial which does not affect the fact that there is an exponential falloff.  The state $|\Psi_m(\nu)\rangle$ is then determined by solution of the inhomogeneous Schr\"odinger equation,
\beq
\left[-\frac{\hbar^2}{2m_e}\nabla^2 -\frac{e^2}r-E(1s)-h\nu\right]\Psi_m(\nu;{\bf r}) = d_m\psi_{1s}({\bf r}).
\label{eq:inhomo}
\eeq
This equation has been extensively studied in the context of the response of a hydrogen atom to electromagnetic radiation (e.g. Ref.~\cite{1928PNAS...14..253P}).  At large $r$, the source falls off exponentially.  Since $E(1s)+h\nu<0$, the operator on the left-hand side of this equation takes the form of a wave equation with imaginary wave number (``$k^2<0$'') at large $r$.  Therefore its solution is exponentially decaying with $r$.  Now we know that as $n\rightarrow\infty$, the wave functions of $|nl\rangle$ states near the origin approach the solution of the Schr\"odinger equation at zero energy:
\beq
\left(-\frac{\hbar^2}{2m_e}\nabla^2 -\frac{e^2}r\right)\psi_{nl}({\bf r})\approx 0 \;\;({\rm small~r}),
\eeq
so that the $\psi_{nl}({\bf r})$'s near the origin are all solutions to the same linear homogeneous differential equation with regular boundary 
conditions at $r=0$.  Therefore they are scaled versions of each other, with the normalization determined by the condition $\int 
|\psi_{nl}|^2\,d^3{\bf r}=1$.  The normalization integral is dominated by regions with $r\gg a_0$ and it is well-known that it enforces 
$\psi_{nl}\propto n^{-3/2}$ in the large $n$ limit.  Given that $\psi_{nl}\propto n^{-3/2}$ near the origin, and that $\Psi_m(\nu;{\bf r})$ is only 
significantly different from zero near the origin, it follows that ${\cal M}_{2\gamma}\propto n^{-3/2}$ and that $d\Gamma_{2\gamma}/d\nu\propto n^{-3}$.

Essentially the same argument applies to helium.  Like the $|1s\rangle$ state of hydrogen, the $|1^1S\rangle$ state of helium is concentrated near the 
origin with an exponential falloff as either electron is moved to large distances from the nucleus.

\begin{figure}
\includegraphics[angle=-90,width=3in]{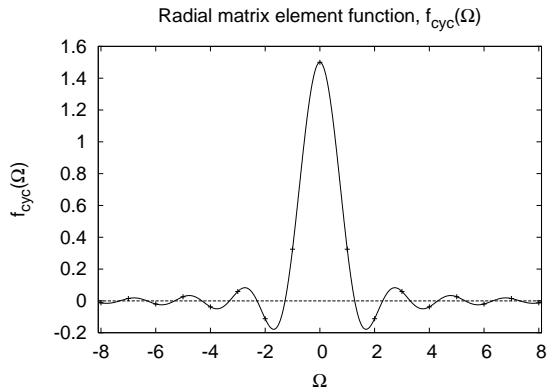}
\caption{\label{fig:f}The function $f(\Omega)$ appearing in Eq.~(\ref{eq:rmatrix-text}), which describes the radial matrix elements for large $n$.  Note that $f(\Omega)$ is largest for small $\Omega$, implying that $\langle n^1L ||{\bf d}||{n'}^1P^o\rangle$ is largest for small $n'-n$.  The points mark integer values of $\Omega$.}
\end{figure}

\begin{figure}
\includegraphics[angle=-90,width=3in]{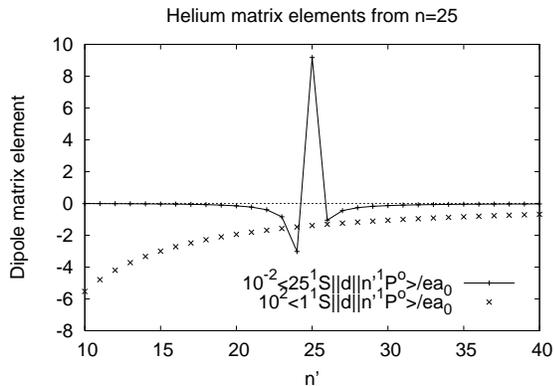}
\caption{\label{fig:n25}The dipole matrix elements $\langle 1^1S||{\bf d}||{n'}^1P^o\rangle$ and $\langle n^1S||{\bf d}||{n'}^1P^o\rangle$ for $n=25$, computed using the formula of Ref.~\cite{1989ApJ...336..504K} (for $1^1S$) and the Coulomb approximation (for $25^1S$).  Note that the matrix elements with $n'$ close to, but not equal to, $n$ are comparable in magnitude to $\langle n^1S||{\bf d}||n^1P^o\rangle$ but have negative instead of positive sign.  Therefore these contributions can interfere destructively with the $n^1P^o$ intermediate level in the matrix element ${\cal M}_{2\gamma}$.}
\end{figure}

Mathematically, the only way to reconcile the ${\cal M}_{2\gamma}\propto n^{-3/2}$ scaling from the above argument with the $\propto n^{1/2}$ scaling obtained by including only the $n'=n$ intermediate state is that there must be a near-exact cancellation of contributions to ${\cal M}_{2\gamma}$.  We can show that this will happen by examining the behavior of the matrix elements for large $n$ and small $s\equiv n'-n$.  This is considered in Appendix~\ref{app:largen}, where it is shown that
\beq
\langle n^1L ||{\bf d}||{n'}^1P^o\rangle \approx (-1)^{L_>}L_>^{1/2}\,(-1)^s ea_0n^2\fcyc(s+\delta_{1L})
\label{eq:rmatrix-text}
\eeq
(c.f. Eq.~\ref{eq:rmatrix}), where $\delta_{1L}$ is the differential quantum defect ($\delta_{10}=0.152$ and $\delta_{12}=0.014$) and $f_{\rm cyc}$ is 
the 
Fourier transform of the cycloid function (Eq.~\ref{eq:f}).  [The existence of an asymptotic limit of the form in Eq.~(\ref{eq:rmatrix-text}) appears to have been noticed by Refs.~\cite{1949RSPTA.242..101B,1968ApJS...16..175O}, and the analytic expression for $\fcyc$ was derived, albeit in a different form, by Ref.~\cite{1981OptSp..51...13D}.]  The function $\fcyc$ is plotted in Fig.~\ref{fig:f}.  What is of note here is that the matrix elements with $s$ equal to a few are of the same order of magnitude as those with $s=0$ ($n'=n$).  Therefore one should include them when obtaining the matrix element ${\cal M}_{2\gamma}$.  Since the matrix element $\langle 1^1S||{\bf d}||{n'}^1P^o\rangle$ scales asymptotically as ${n'}^{-3/2}$, and the energy $\Delta E({n'}^1P^o)$ approaches a constant at large $n'$, these can be considered constant for $|s|\ll n$.  We may thus include the values with $s_{\rm min}\le s\le s_{\rm max}$ by writing
\beqa
{\cal M}_{2\gamma} \!\! &\approx& \!\! \frac{en^2}{a_0}
\langle 1^1S || {\bf d} || n^1P^o \rangle
(-1)^{L_>}L_>^{1/2}
\nonumber \\ && \!\! \times
\sum_{s=s_{\rm min}}^{s_{\rm max}} (-1)^s\fcyc(s+\delta_{1L})
\nonumber \\ && \!\! \times
\Biggl(\frac 1{\Delta E(n^1P^o)-h\nu}
\nonumber \\ &&
 + \frac 1{\Delta E(n^1P^o)-h\nu'}\Biggr).
\eeqa
The first line in this equation scales as $n^{1/2}$, which when squared gives the DG05 scaling $d\Lambda_{2\gamma}/d\nu\propto n$.  One must be mindful of the second line however, which modifies the prefactor of $n^{1/2}$ in the asymptotic scaling of ${\cal M}_{2\gamma}$.  One would expect to get a better estimate of the asymptotic scaling by taking the limits $s_{\rm min}\rightarrow-\infty$ and $s_{\rm max}\rightarrow\infty$.  However we show in Appendix~\ref{app:largen} that (see Eq.~\ref{eq:0})
\beq
\sum_{s=-\infty}^\infty (-1)^s\fcyc(s+\delta_{1L}) = 0.
\label{eq:0-text}
\eeq
Therefore for large $n$ one expects the contribution to ${\cal M}_{2\gamma}$ from states with $n'$ near $n$ to be $n^{1/2}$ times something approaching 
$0$.  This of course implies that one cannot find the large-$n$ behavior of ${\cal M}_{2\gamma}$ by the DG05 argument (except possibly by considering 
higher-order corrections to Eq.~\ref{eq:rmatrix-text}) -- one can only say that it scales slower than $n^{1/2}$.  The near-cancellation is illustrated 
graphically in Fig.~\ref{fig:n25} for the $25^1S\rightarrow 1^1S$ decay.

In summary, we conclude that (i) the actual large-$n$ behavior of the two-photon decay rate is $d\Lambda/d\nu\propto n^{-3}$, and (ii) the apparent discrepancy between this and DG05 is due to a cancellation of the matrix elements as summarized by Eq.~(\ref{eq:0-text}).

\subsection{Rate estimates}
\label{ss:newrate}

\begin{figure}
\includegraphics[angle=-90,width=3in]{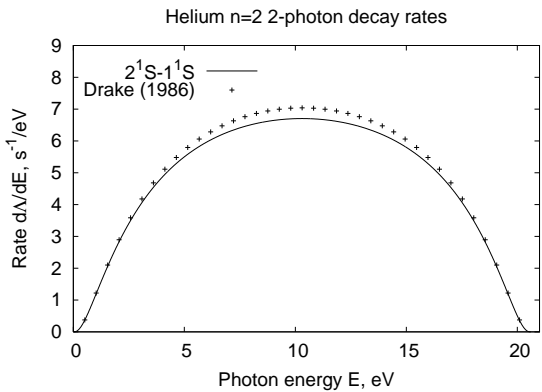}
\caption{\label{fig:he2}The 2-photon spectrum from the $2^1S$ level of \HeI.  Note that there are no resonances in the spectrum since there are no energetically allowed 1+1 electric dipole decays from $2^1S$.  The points are the more detailed calculations by Drake \cite{1986PhRvA..34.2871D}.}
\end{figure}

We have estimated the two-photon transition rates for small $n$ by direct summation of the matrix element products in Eq.~(\ref{eq:m}).  We note that such a calculation does not require a detailed re-analysis of the atomic physics, at least to a first approximation, because all one needs to know are the energies and dipole matrix elements for the $^1S$, $^1P^o$, and $^1D$ levels, which have already been calculated.  (The exceptions are the continuum levels, for which the information available is more sparse, nevertheless as we argue below there are detailed calculations for the first several eV of the continuum, which are dominant.)  We have obtained these as follows:
\newcounter{RE1}
\begin{list}{\arabic{RE1}. }{\usecounter{RE1}}
\item For the $n^1S$--${n'}^1P^o$ with $n,n'\le 9$, we use the oscillator strengths from Ref.~\cite{1984PhRvA..29.2981K}.  
\item The $1^1S-10^1P^o$ oscillator strength is from Ref.~\cite{1988ApJ...329..493K}.  For $1^1S$--${n'}^1P^o$ transitions with $n'>10$, we have used the asymptotic formula for the oscillator strength from Ref.~\cite{1989ApJ...336..504K}.
\item For the $n^1S$--${n'}^1P^o$ transitions with $n,n'\ge 10$, we used the Coulomb approximation \cite{1949RSPTA.242..101B}.
\item For transitions from $n^1S$ to the $^1P^o$ continuum, we used the TOPBase photoionization cross sections \cite{1993A&A...275L...5C} converted to matrix elements in accordance with
\beq
\sigma = \frac{4\alpha}{3e^2} h\nu \frac{dn'}{dE} \left| \langle n^1S || {\bf d} || {n'}^1P^o\rangle \right|^2,
\label{eq:sigma}
\eeq
where $dn'/dE$ is the density of continuum states.  This turns the continuum contribution to Eq.~(\ref{eq:m}) into an integral over energy.
\item Dipole matrix elements for $S$--$P^o$ transitions are obtained from the standard formula,
\beq
\left|\langle n^1S || {\bf d} || {n'}^1P^o \rangle\right|^2 = \frac{3e^2a_0^2{\cal R}f_{n^1S\rightarrow {n'}^1P^o}}{E({n'}^1P^o)-E(n^1S)}.
\eeq
The helium atom wavefunctions for $m=0$ are all real and hence $\langle n^1S || {\bf d} || {n'}^1P^o \rangle$ is purely real, however a sign ambiguity exists.  We have taken the sign for $S$--$P^o$ to be negative for $n\neq n'$ and positive for $n=n'$, as this is what is found using the Coulomb approximation or hydrogenic wavefunctions.
\item The $P^o$--$D$ dipole matrix elements are taken to be hydrogenic.
\end{list}

Note that this approach is expected to break down for continuum $^1P^o$ levels with very large energies.  In particular the continuum $^1P^o$ wave 
functions become less hydrogenic at higher energies where the outer electron penetrates deeper into the He$^+$ ``core,'' and it becomes very 
non-hydrogenic as one approaches the double-excitation resonance region 60 eV above the ground state.  At still higher energies there are multiple 
continua, so it is no longer valid to compute bound-free matrix elements using Eq.~(\ref{eq:sigma}) -- the matrix elements actually contain information 
that is not contained in the cross section.  Fortunately, these subtleties have little effect at the level of accuracy required here: we find that 
neglecting continuum levels with energies more than $0.5{\cal R}$ (6.8 eV) above threshold makes at most a change of 30\% ($3^1S\rightarrow 1^1S$) or 
1\% ($3^1D\rightarrow 1^1S$) to $d\Lambda/d\nu$, except in the immediate vicinity of the nulls (whose positions are slightly shifted).  Since we will 
find that the total correction to the recombination history due to nonresonant two-photon decays is $\sim 4\times 10^{-4}$, we believe 
that our basic conclusion that nonresonant two-photon decays are unimportant is robust even if the rate estimates are off by several tens of percents.

We show the two-photon decay rates we obtain for the \HeI\ $2^1S$ level in Fig.~\ref{fig:he2}.  The total decay rate we obtain is $49\,$s$^{-1}$, in 
comparison with the more detailed atomic physics calculations, which give $51\,$s$^{-1}$ \cite{1986PhRvA..34.2871D}.  This provides a check on the 
accuracy of our method.

\begin{figure}
\includegraphics[angle=-90,width=3in]{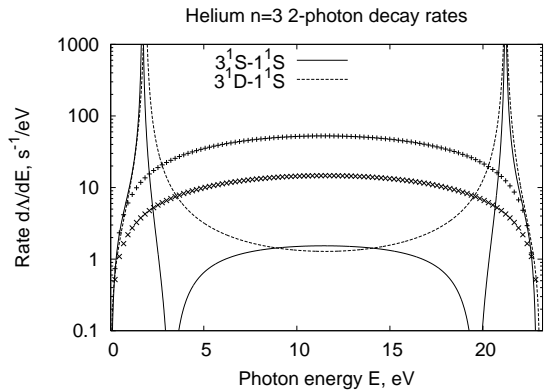}
\caption{\label{fig:he3}The 2-photon spectrum from the $3^1S$ and $3^1D$ levels of \HeI.  Note the resonance at $21.2\,$eV corresponding to the 
$2^1P^o$ intermediate level.  There are also resonances at much lower energies corresponding to the optical transitions in \HeI, 
$3^1S$--$2^1P^o$ and $3^1D$--$2^1P^o$.  Also note the nulls in the two-photon rate from $3^1S$.  The series of points are the results considering only 
the $n'=n$ term in the matrix element, Eq.~(\ref{eq:m}), for $3^1S$ (upper series) and $3^1D$ (lower series).  Note that keeping only this term is a 
poor approximation except at the very ends of the spectrum.}
\end{figure}

\begin{figure}
\includegraphics[angle=-90,width=3in]{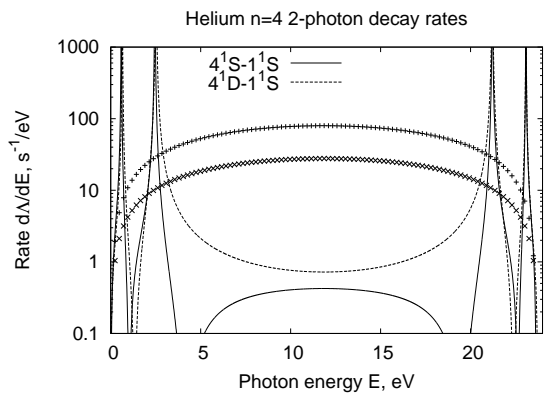}
\caption{\label{fig:he4}The 2-photon spectrum from the $4^1S$ and $4^1D$ levels of \HeI.  There are two pairs of resonances corresponding to the 1+1 decays via the $2,3^1P^o$ intermediate levels.  The series of points are the results considering only the $n'=n$ term in the matrix element, Eq.~(\ref{eq:m}), for $4^1S$ (upper series) and $4^1D$ (lower series).}
\end{figure}

Also of interest are the two-photon rates from the $n=3$ (Fig.~\ref{fig:he3}) and $n=4$ (Fig.~\ref{fig:he4}) levels.  These two-photon spectra show resonances at the positions of allowed 1+1 transitions.  We have shown the results using only the $n'=n$ terms as a series of points in each plot; one can see that this is a poor approximation across most of the spectrum.  In particular, far from the resonances, this substantially overestimates the rate because it neglects destructive interference between different intermediate levels.

\section{Effect of nonresonant two-photon transitions}
\label{sec:nonres}

Now that we have obtained the two-photon rates, we would like to understand how much \HeI\ recombination is modified by including them.  The main 
contribution comes from the lower values of $n$, both because of their faster rates and because the lower-$n$ states have higher occupation 
probabilities.  This section considers the addition of {\em nonresonant} two-photon decays from $3\le n\le 5$ and nonresonant Raman scattering from 
$2\le n\le 5$, and finds a negligible effect.

There is one subtlety involved in including higher-order two-photon transitions, which was recognized already in DG05.  It is the existence of the $1+1$ resonances, which cause the two-photon rate to be very large when the photons are emitted in allowed electric dipole lines.  Photons emitted in these lines (i) have a high probability of being re-absorbed, and (ii) are in any case already included in the treatment of Paper I, which included all of the one-photon transitions.  In this paper, we will handle this issue by dividing the two-photon spectrum into nonresonant and resonant pieces, which are treated separately.  Here ``nonresonant'' simply means that the emitted photons are detuned from the 1+1 resonance by some minimum frequency offset $\nucut$.  The idea is to show in this section that the nonresonant transitions have no significant effect on \HeI\ recombination, and then in the next section consider whether the approximations made in Paper I about resonant two-photon transitions are valid.  The offset $\nucut$ is arbitrary and was chosen so as to make both the arguments in this section and the following section valid.  Precisely the same subtlety arises in considering Raman scattering, which has resonances such as $2^1S\rightarrow 2^1P^o\rightarrow 1^1S$, and we handle the problem in precisely the same way.  The choice of the frequency offset that we use is $0.14n^{-3}{\cal R}=460n^{-3}\,$THz for the offset from the \HeI\ $n^1P^o$--$1^1S$ line; the motivation is that we do not want our definition of ``resonant'' photons to overlap with the intercombination line \HeI] $n^3P^o$--$1^1S$. (There is an overlap with the quadrupole lines [\HeI] $n^1D$--$1^1S$, however as we argue in Paper III, these lines do not matter anyway.)

In the absence of the spectral distortion, nonresonant two-photon transitions and Raman scatterings can be trivially included in a level code as an 
additional rate,
\beqa
\dot x|_{n^1L\rightarrow 1^1S} \!\!&=&\!\! [\Gamma_{2\gamma}(n^1L)+\Gamma_{\rm Raman}(n^1L)]
\nonumber\\&&\!\!\times
\left(x_{n^1L} - \frac{g_{n^1L}}{g_{1^1S}}x_{1^1S} e^{-E_{n^1L}/\kB\Tr}\right),\;\;
\label{eq:include}
\eeqa
where the term with $x_{1^1S}$ accounts for thermal re-excitations of ground-state helium atoms determined via the principle of detailed balance.  The two-photon and Raman scattering rates, $\Gamma_{2\gamma}(n^1L)$ and $\Gamma_{\rm Raman}(n^1L)$, are obtained by integration of Eqs.~(\ref{eq:g2g}) and (\ref{eq:graman}) with blackbody radiation profiles, except that regions in the integral where the higher-energy photon lies within $\nucut$ of an allowed resonance are excluded.  We have obtained fitting formulas for the two-photon rates, which are given in Table~\ref{tab:fit}.  We have also included nonresonant Raman scattering from the $2^1S$ level, which is well fitted by
\beq
\Gamma_{\rm Raman}(2^1S) = 12.8t^{1.5}e^{-2.125/t}\,{\rm s}^{-1},
\label{eq:raman-2s}
\eeq
where $t=T_r/4000\,$K is in the range $1\le t\le 2$.  For $1\le t\le 2$, Eq.~(\ref{eq:raman-2s}) and the formulas in Table~\ref{tab:fit} agree with our numerical calculations to within 1\%, which is probably better than the accuracy of our rates.

\begin{table}
\caption{\label{tab:fit}Fitting formulas for the two-photon rates $\Gamma_{2\gamma}(n^1L)+\Gamma_{\rm Raman}(n^1L)$ appearing in Eq.~(\ref{eq:include}) for the $n=3$, 4, and 5 levels of helium.  The rate in units of s$^{-1}$ is written as $a+bt$, where $t=T_r/4000\,$K.  The formulas are valid in the range $1<t<2$ of interest to helium recombination.  Note that these numbers include only the nonresonant contribution, defined as having the emitted photon at least $0.14{n'}^{-3}{\cal R}$ from the \HeI\ ${n'}^1P^o$--$1^1S$ line.}
\begin{tabular}{clrlrcclrlr}
\hline\hline
Upper level & & $a$ & & $b$ & &
Upper level & & $a$ & & $b$ \\
\hline
$3^1S$ & & 20.5 & & 16.0 & &
$3^1D$ & & 94.0 & &  8.0 \\
$4^1S$ & & 10.2 & & 12.5 & &
$4^1D$ & & 42.3 & &  6.0 \\
$5^1S$ & &  6.1 & &  9.2 & &
$5^1D$ & & 21.5 & &  5.0 \\
\hline\hline
\end{tabular}
\end{table}

\begin{figure}
\includegraphics[angle=-90,width=3.2in]{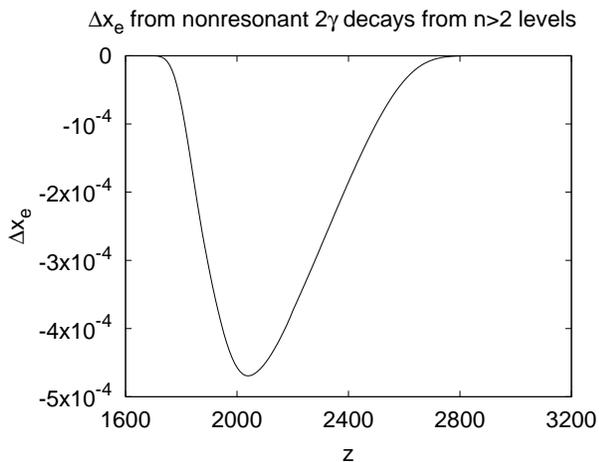}
\caption{\label{fig:nr_contrib} The contribution of the non-resonant part of the two photon rate in \HeI\ for $n>2$ produces a maximum change in the 
free 
electron fraction of several$\times 10^{-4}$.}
\end{figure}

Fig.~\ref{fig:nr_contrib} shows the change in the electron abundance due to nonresonant two-photon transitions.  The effect is at the level 
of a few times $10^{-4}$ and can be neglected.

\section{The $1+1$ resonances and finite linewidth}
\label{sec:res}

In Sec.~\ref{sec:nonres}, we considered the influence of the nonresonant two-photon transitions on \HeI\ recombination.  We know however that the total 
two-photon transition rate is dominated by the $1+1$ resonances (except in the case of the $2^1S$ level, which has no such resonances).  If this 
additional rate is na\"ively added to the recombination equations in the manner of Eq.~(\ref{eq:include}), \HeI\ recombination becomes described by the 
Saha equation.  However we know that this na\"ive addition is incorrect because photons emitted within resonance lines with lower level $1^1S$ will 
likely be re-absorbed.  In order to understand the effect of resonant two-photon transitions, we must understand the transport of radiation within the 
\HeI\ $n^1P^o$--$1^1S$ lines.  We presented a simplified analysis of this in Paper I, where photons were injected into the line by resonant two-photon 
emission and \HI\ recombination, transported by coherent (Rayleigh) scattering and Hubble redshifting, and finally removed by resonant two-photon 
absorption and \HI\ photoionization.  The analysis in Paper I makes the approximation that the \HeI\ line is infinitesimally thin relative to variation in the 
radiation phase space density and phase space factors.  The purpose of this section is 
to test the validity of these assumptions in certain special cases and understand the errors introduced.  The basic method here is to reconsider the 
$2^1P^o$--$1^1S$ line including the deviation from Voigt profile in the far damping wings, and including the deviation of the radiation profile from 
steady state.  We incorporate these corrections into the level code and show that the modification to recombination is small ($|\Delta x_e|\sim 3\times 10^{-4}$).

The specific assumptions made in Paper I that we would like to test are:
\newcounter{PIA}
\begin{list}{\arabic{PIA}.}{\usecounter{PIA}}
\item\label{it:simple} The two-photon emission profile can be described by a Voigt distribution, i.e. we neglected the possible interference with neighboring 1+1 resonances, and the variation of the photon phase space factor $\nu^3{\nu'}^3$ and the phase space density factor $[1+{\cal N}(\nu)][1+{\cal N}(\nu')]$ (cf. Eq.~\ref{eq:rate-2g}) across the line width.  (A similar assumption applies to our treatment of Raman scattering and two-photon absorption.)
\item The \HeI\ line was treated as being in steady state, i.e. we assumed that the rate of injection of photons equaled the loss rate.  In reality, there are always a few photons within the line, and as this number of photons increases (or decreases) there is a corresponding speed-up (or slow-down) of \HeI\ recombination.
\end{list}

We will examine these assumptions here in the context of the $2^1P^o$--$1^1S$ line, which was found in Paper I to produce the most important effect.   To simplify the calculation, we will also assume when calculating line shapes that the excited levels in \HeI\ are in equilibrium. (This was found in Paper I to be a good approximation and is described quantitatively in Paper III.)  We will introduce the notation $\nu_-$ and $\nu_+$ to denote the minimum and maximum frequencies of the resonance, i.e.
\beq
\nu_\pm = \nu_{1^1S-n^1P^o} \pm 0.14n^{-3}{\cal R},
\eeq
where here $n=2$.

\subsection{Finite linewidth}
\label{ss:finitediscussion}

The calculations involving transport and incoherent scattering in Paper I made use of an approximate symmetry of the thermal radiation field in the 
neighborhood of the line.  Here, so long as the linewidth is negligible compared to $\kB\Tr/h$, differences in the thermal radiation field on either 
side of the line can be neglected.  In this section we will elaborate on this and argue that: 1) a linewidth of much less than $\kB\Tr/h$ means that a 
photons are just as likely to be absorbed on either the red or the blue side of the line, and 2) that introducing a finite width to the line means that 
photons will be scattered differentially depending on whether they are on the red or the blue side of the line.  This will result in photons being 
``pumped'' redward by the asymmetry in the thermal radiation field across the line.

Incoherent scattering is inherently a multi-photon process.  Here, a photon is scattered and another is re-emitted and distributed over the line's 
profile, with no memory of the incoming energy.  The only way for this reaction to proceed is for some number of other particles to recoup the change 
in energy (as compared to coherent scattering, where the photon's energy is exactly conserved in the atom's rest frame).

Consider the case of incoherent scattering off \HeI\ $1^1S$ through $2^1P^o$ with an excursion to $3^1D$.  Here the incoming photon (A) in the \HeI\ 
$\lambda 584$ line excites the atom to the $2^1P^o$ level.  This excited atom resonantly scatters a second photon (B$\rightarrow$C) in the \HeI\ 
$\lambda 6678$ line via the $3^1D$ resonance.  Finally the atom decays back to the ground state, emitting a photon in the $\lambda 584$ line (D).  
This process can be viewed as a resonant two-photon absorption of photons A and B, followed by two-photon decay emitting C and D.

In principle photon B could be any photon drawn from the blackbody radiation field, however because of the narrow $3^1D$ resonance in helium, the 
photon absorbed will almost always have energy $E_{\rm B}=\Delta E(3^1D)-E_{\rm A}$.  The excited \HeI\ $3^1D$ atom then undergoes two-photon decay to 
the ground state via the $2^1P^o$ intermediate level (i.e. it emits photons C and D).  To a very good approximation, the energy distribution of D is 
independent of $E_{\rm A}$ -- hence the term ``complete redistribution'' -- and in the vicinity of resonance it has the form of a Lorentz profile (or a 
Voigt profile in the comoving instead of atom frame).

\begin{figure}
\includegraphics[width=3.2in]{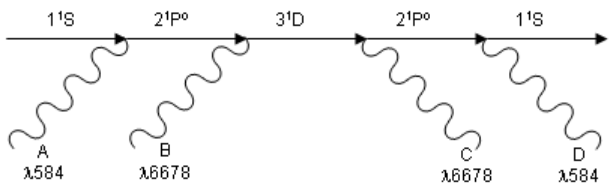}
\caption{\label{fig:incscat}An example of an incoherent scattering process in \HeI\ considered in Sec.~\ref{ss:finitediscussion}.}
\end{figure}

Now, suppose that the atom absorbs photon A on the blue side of the $\lambda 584$ line.  Then it can be absorbed in combination with a photon B on the red side of $\lambda 6678$, whereas if photon A is on the red side of the $\lambda 584$ line then it requires B to be on the blue side of $\lambda 6678$.  Since in a blackbody distribution for photon B there are more photons on the red side of the line, this means that there is an enhancement in the cross section for absorbing photon A from the blue side of $\lambda 584$, and a suppression for absorbing it from the red side of $\lambda 584$.  This means that (even in the absence of Hubble redshifting) $\lambda 584$ photons spend on average more time on the red side of the line, so that $\pha$ is greater there.  

The same conclusion could also have been reached by a thermodynamic argument: since incoherent scattering changes the energy of the $\lambda 584$ line photons by exchanging their energy with that of the $\lambda 6678$ photons (and with other low-energy photons if we consider the other lines connecting \HeI\ $2^1P^o$ to other excited levels), and the radiation in these lines is essentially blackbody, it follows that photons near the $\lambda 584$ line will then be driven toward a Bose-Einstein distribution with temperature $\Tr$ and some chemical potential determined by the total number of such photons.  Since $\pha\ll 1$, this is equivalent to a Boltzmann distribution, $\pha\propto e^{-h\Delta\nu/\kB\Tr}$.  We will see this behavior mathematically from Eq.~(\ref{eq:ssl}).  (Note that whether $\pha\propto e^{-h\Delta\nu/\kB\Tr}$ is actually achieved depends on whether incoherent scattering can operate efficiently before Hubble redshifting moves the photons out of the line, a question that can only be settled by solving the equations.)

\begin{figure}
\includegraphics[width=3.2in]{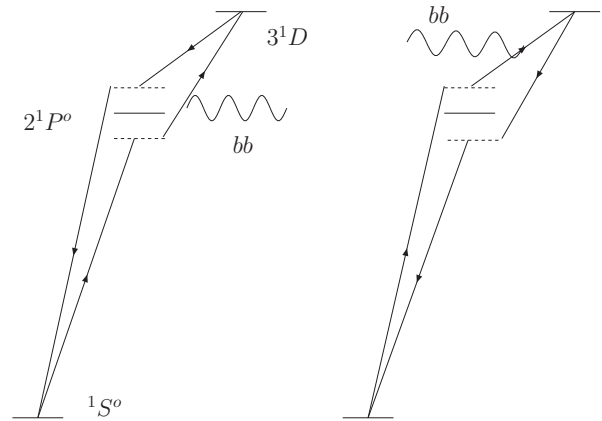}
\caption{\label{fig:finitelinewidth}Representation of complete redistribution as a two photon process, with one photon from the thermal distribution.  
In the left frame, a photon is absorbed on the red side of the line, then assisted to $3^1D$ by a black-body photon.  In the right frame, a photon is absorbed on the blue side of the line and assisted by a lower energy black-body photon.  The virtual levels have energies offset from $E(2^1P^o)$; the fractional difference between the forward and backward scattering rates is of order the frequency difference times $h/\kB\Tr$.  Because there are more low-energy thermal photons, scattering to the blue side of the line is slightly preferred.}
\end{figure}

In summary, we have argued that incoherent scattering through a finite linewidth enhances the phase space density on the red side of the $\lambda 584$ line.  In the limit that the width of a line is taken to be negligible compared to $\kB\Tr/h$, incoherent scattering redistributes photons and pushes the radiation phase space density near line center to some constant $\pha_L$, in equilibrium with the line.  This flattening tendency is implicit in the analysis of Paper I, and has been noted several times in the recombination literature \cite{1989ApJ...338..594K, 1990ApJ...353...21K}.  

Viewed in this way, incoherent scattering is the sum of two-photon scattering processes for which the excited level is an intermediate state 
(resonance) in the full two-photon rate.  The goal, then, is to consider the full expression for all two photon processes, and separate the transport 
physics around one of the intermediate state resonances, e.g. $2^1P^o$.  It is possible, then, to write an effective one-photon transport equation 
(where the other photon is drawn from the black body) for incoherent scattering to this intermediate state.  Aside from the change in the line profile, 
the finite linewidth introduces two new pieces of physics: the tendency to drive the radiation spectrum to $\pha\propto e^{-h\nu/\kB\Tr}$ instead of a 
constant, and the fact that the line is not exactly in steady state (i.e. there is a $\partial\pha/\partial t$ term in the transport equation).  This 
section considers these issues and introduces a crude correction to the rate equations.  We incorporate this in the level code and find only a small 
correction (a few times $10^{-4}$).  This correction is not included in the final version of the recombination history presented in Paper III.

\subsection{Line transport with complete redistribution and no \HI\ opacity}
\label{ss:linetransnoopacity}

The case we consider here is that where the \HI\ opacity within the line and frequency diffusion due to Doppler shift in repeated resonant scatterings 
can be neglected.  This is useful for testing assumption \#\ref{it:simple} on our list (Sec.~\ref{sec:res}).  The assumption of negligible \HI\ opacity is valid in the 
early stages of \HeI\ recombination, i.e. $2200<z<2800$.  The frequency diffusion was included in Paper I and neglecting it was found to introduce no 
significant error: $|\Delta x_e| < 2 \times 10^{-4}$. 

The resonances in consideration are optically thick and the radiation rapidly approaches equilibrium around their line centers.  Because of this, Doppler broadening can be neglected to a good approximation and we can consider the radiative processes as occurring in the atom's rest frame.
The differential equation describing the radiation field is
\beqa
\frac{\partial\pha}{\partial t} \!\!&=&\!\!
H\nu\frac{\partial\pha}{\partial\nu} + \frac{c^3\nH}{8\pi\nu^2}\sum_i x_i\frac{d\Gamma_i}{d\nu} 
\nonumber \\ &&\!\!
- \nH c\sum_i\sigma_i^{(2\gamma+\rm Raman)}(\nu) \pha,
\label{eq:l1}
\eeqa
where the sum is over excited levels of \HeI\ that can undergo two-photon decay or Raman scattering to the ground state, $d\Gamma_i/d\nu$ is their rate of producing line photons per unit frequency, and $\sigma_i^{(2\gamma+\rm Raman)}$ is the cross section for removing line photons via two-photon absorption or Raman scattering to level $i$.  This is a strong function of $\nu$, but we will drop the explicit argument to stay concise.  (Though officially a three-body process, it is possible to define a cross section for two-photon absorption of a $2^1P^o$--$1^1S$ line photon since the other photon comes from much lower energies where the CMB can be treated as a blackbody.)  By detailed balancing of the level $i$ contributions to the second and third terms on the right hand side, we find
\beq
\frac{c^3\nH}{8\pi\nu^2} \frac{d\Gamma_i}{d\nu} \frac{g_i}{g_{1^1S}}e^{-\Delta E_i/\kB\Tr}
= \nH c\sigma_i^{(2\gamma+\rm Raman)} e^{-h\nu/\kB\Tr},
\label{eq:l2}
\eeq
which allows us to derive the cross sections for two-photon absorption to each level.  Then since the excited levels are in equilibrium with each other we have
\beq
x_i = \frac{g_i}{g_{2^1P^o}}x_{2^1P^o}e^{-[E_i-E(2^1P^o)]/\kB\Tr}.
\label{eq:l3}
\eeq
Combining Eqs.~(\ref{eq:l1}), (\ref{eq:l2}), and (\ref{eq:l3}), we get
\beqa
\frac{\partial\pha}{\partial t}
 \!\!&=&\!\! 
H\nu\frac{\partial\pha}{\partial\nu}
\nonumber \\ &&\!\!
-\nH cx_{1^1S}\sigma^{(2\gamma+\rm Raman)}
\nonumber \\ &&\!\!\times
\left( \pha - \frac{x_{2^1P^o}}{3x_{1^1S}}e^{-h\Delta\nu/\kB\Tr}\right).
\label{eq:ssl}
\eeqa

There are several approaches available for solving Eq.~(\ref{eq:ssl}).  We will take an approach that allows us to separate the effects of the line profile from the steady state approximation.  The method is to multiply the left hand side of Eq.~(\ref{eq:ssl}) by an artificial expansion parameter $\epsilon$, which will eventually be taken to equal $1$.  We may then expand
\beq
\pha = \pha_0 + \epsilon\pha_1 + \epsilon^2\pha_2 + ...;
\eeq
equating coefficients of $\epsilon^j$ in Eq.~(\ref{eq:ssl}) then leads to the following situation.  For $j=0$, we find
\beqa
0 \!\!&=&\!\! 
H\nu\frac{\partial\pha_0}{\partial\nu}
\nonumber \\ &&\!\!
-\nH cx_{1^1S}\sigma^{(2\gamma+\rm Raman)}
\nonumber \\ &&\!\!\times
\left( \pha_0 - \phaL e^{-h\Delta\nu/\kB\Tr}\right),
\label{eq:e0}
\eeqa
where $\phaL=x_{2^1P^o}/3x_{1^1S}$.  That is, $\pha_0$ satisfies the steady-state equation.  The higher-order terms satisfy
\beq
\frac{\partial\pha_{j-1}}{\partial t} = H\nu\frac{\partial\pha_j}{\partial\nu}
- \nH cx_{1^1S}\sigma^{(2\gamma+\rm Raman)}\pha_j
\label{eq:e1}
\eeq
for $j\ge 1$.  Since photons enter from the blue side of the line, the boundary condition $\pha(\nu_+)$ is satisfied; the Taylor expansion of this condition in $\epsilon$ is that $\pha_0(\nu_+)=\pha(\nu_+)$, and $\pha_j(\nu_+)=0$ for $j\ge1$.  We may think of the $\pha_j$ for $j\ge1$ as successive corrections to the steady-state solution.  For each $j$, a numerical solution may be obtained by starting at $\nu=\nu_+$ and using a stiff ODE integrator in the redward direction until we reach $\nu_-$.

In order to translate our results for the line profile into effects on recombination, we need two numbers.  One of these is the photon phase space density $\pha(\nu_-)$ emerging from the red side of the line, necessary to compute feedback.  The other is the net decay rate to the ground state, which is obtained by subtracting the downward from the upward rates:
\beqa
\dot x_\downarrow \!\!&=&\!\! \int_{\nu_-}^{\nu_+} \Biggl[\sum_i x_i\frac{d\Gamma_i}{d\nu}
\nonumber \\ &&
- \frac{8\pi\nu^2}{c^2}x_{1^1S}\sum_i\sigma_i^{(2\gamma+\rm Raman)}\pha
\Biggr]\,d\nu.
\label{eq:bad}
\eeqa
The downard and upward rates nearly cancel, so numerically the best way to compute this is not to evaluate Eq.~(\ref{eq:bad}) directly from the solution, but rather to use Eq.~(\ref{eq:l1}) to re-write it as
\beq
\dot x_\downarrow = -\int_{\nu_-}^{\nu_+} \frac{8\pi\nu^2}{\nH c^3}\left(H\nu\frac{\partial\pha}{\partial\nu}
-\epsilon\frac{\partial\pha}{\partial t}\right)\,d\nu.
\label{eq:good}
\eeq
The steady-state solution is obtained in Eq.~(\ref{eq:good}) by setting $\pha=\pha_0$ and $\epsilon=0$ (i.e. dropping the time derivative term).  The first-order solution in $\epsilon$ is
\beqa
\dot x_\downarrow \!\!&=&\!\! -\int_{\nu_-}^{\nu_+} \frac{8\pi\nu^2}{\nH c^3}\left[ H\nu\frac{\partial(\pha_0+\pha_1)}{\partial\nu}
-\frac{\partial\pha_0}{\partial t}\right]\,d\nu.\;\;
\label{eq:first}
\eeqa

\begin{figure}
\includegraphics[angle=-90,width=3.2in]{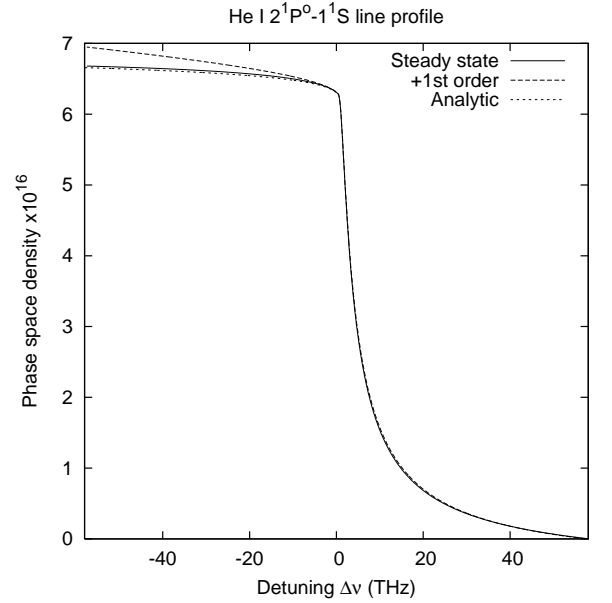}
\caption{\label{fig:nss}The phase space density in the $2^1P^o$--$1^1S$ line using the parameters $z=2175$, $x_{\rm HeI}=0.02825$, $\phaL=6.307\times 10^{-16}$, $\pha(\nu_+)=5.277\times 10^{-19}$, $\dot x_{\rm HeI}=0.1596H$, $\dot\phaL=-2.444\times 10^{-15}H$, and $\dot\pha(\nu_+)=-4.073\times 10^{-17}H$.  These parameters occurred during the first feedback iteration of the recombination history.  The solid line shows the steady-state solution $\pha_0$, while the long-dashed line is the first-order correction $\pha_0+\pha_1$.  The short-dashed line is the analytic approximation to the steady-state solution from Eq.~(\ref{eq:steadystate}); note that it is plotted only for $\Delta\nu<0$.}
\end{figure}

The line profile is shown in Fig.~\ref{fig:nss} for a typical set of parameters, and is compared with the infinitesimal linewidth approximation, the steady state solution, and the analytic model of Appendix~\ref{app:ss}.  The most important property of the solution, which is generic, is that $\pha>\phaL$ for $\Delta\nu<0$.  That is, the effect of using the full $\pha_0+\pha_1$ in Eq.~(\ref{eq:first}) instead of just a step function at the line is to enhance the decay rate and accelerate recombination.  On the other hand, $\partial\pha_0/\partial t<0$, so the correction due to the line not being exactly in steady state is of the opposite sign: it delays recombination.

\subsection{Inclusion in the level code and the effect on recombination}

The basic strategy in including the finite linewidth effects in the level code is to determine the corrections to the phase space density $\pha(\nu_-)$ on the red side of the line and the net downward transition rate $\dot x_\downarrow$.  This section describes how we do this, and the results when the correction is incorporated in the level code.

The recombination level code depends on $\pha(\nu_-)$ and the reaction rates implied by finite linewidth.  In general these depend on the parameters $\{ z, x_{\rm HeI}, \dot x_{\rm HeI}, \phaL, \dot \phaL, \pha(\nu_+), \dot \pha(\nu_+) \}$.  Since the equation for $\pha$ is linear, $\pha_0$ depends linearly on the parameters $\{\phaL,\pha(\nu_+)\}$, and $\pha_1$ depends linearly on the parameters $\{\phaL, \dot \phaL, \pha(\nu_+), \dot \pha(\nu_+)\}$.  Also $\dot x_{\rm HeI}$ enters only via $\dot\pha_0$, which is the source for $\pha_1$ (c.f. Eq.~\ref{eq:e1}).  From this one can see that the phase space density may be written as
\beqa
\pha(\nu) \!\!&=&\!\! c_0(\nu)\phaL + c_1(\nu)\dot\phaL \nonumber \\   &&\!\!
+ c_2(\nu)\pha(\nu_+) + c_3(\nu)\dot\pha(\nu_+) \nonumber \\ &&\!\!
+ \dot x_{\rm HeI}\left[ c_4(\nu)\phaL + c_5(\nu)\pha(\nu_+) \right],
\eeqa
where the $c_i(\nu)$ depend on $z$, $x_{\rm HeI}$, and cosmological parameters.  Thus if we want $\pha(\nu_-)$, then for each cosmology an interpolation grid can be constructed to give $c_i(\nu_-)$ in terms of the independent variables $z$ and $x_{\rm HeI}$.  A similar result holds for $\dot x_\downarrow$ since it is a linear function of $\pha_0$, $\dot\pha_0$, and $\pha_1$.

The easiest way to incorporate the new effect in the level code is actually to calculate the {\em correction} to $\pha(\nu_-)$ and $\dot x_\downarrow$.  
In the case of infinitesimal linewidth, no continuum opacity, and high optical depth (literally, negligible probability of a photon redshifting through 
the line without undergoing an incoherent scattering -- see Appendix~\appprob\ of Paper I), we have $\Pesc=\tauS^{-1}$.  In this case, the photon phase 
space density on the red side of the line is $\phaL$ and the downward transition rate is $8\pi H\nu_{\rm line}^3(\phaL-\pha_+)/\nH c^3$.  
If we ask about the photon phase space density at $\nu_-<\nu_{\rm line}$, and specify the incoming (blue-side) phase 
space density at $\nu_+>\nu_{\rm line}$, this becomes
\beqa
\pha(\nu_-,z_-) &\rightarrow& \phaL(z) {\rm ~~~and} \nonumber \\
\dot x_\downarrow(z) &\rightarrow& \frac{8\pi H\nu_{\rm line}^3}{\nH c^3}[\phaL(z) - \pha(\nu_+,z_+)],
\eeqa
where
\beq
1+z_\pm = \frac{\nu_\pm}{\nu_{\rm line}}(1+z)
\eeq
since it takes a finite amount of time for photons to redshift through the line.  One may thus define a ``finite linewidth correction''
\beq
\delta\pha(\nu_-,z_-) \equiv \pha(\nu_-,z_-) - \phaL(z)
\label{eq:interp1}
\eeq
for the phase space density on the red side of the line (used for feedback), and a similar correction
\beq
\delta\dot x_\downarrow(z)\equiv \dot x_\downarrow(z) -
\frac{8\pi H\nu_{\rm line}^3}{\nH c^3}[\phaL(z) - \pha(\nu_+,z_+)]
\label{eq:interp2}
\eeq
for the transition rate.

\begin{figure}
\includegraphics[angle=-90,width=3.2in]{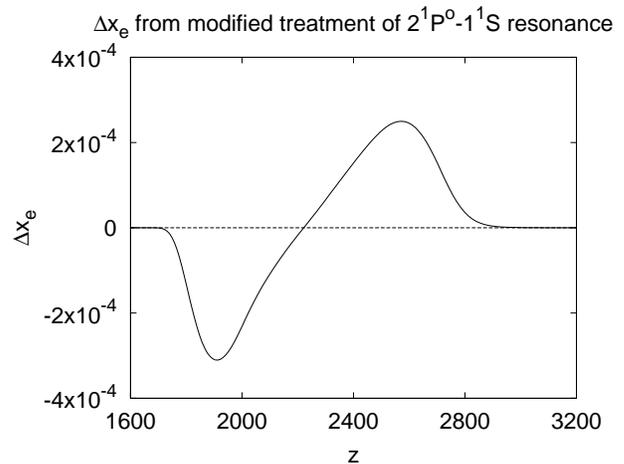}
\caption{\label{fig:diff-fl}The change in the recombination history from the modified treatment of the $2^1P^o$--$1^1S$ resonance.  Note that the 
effect on the electron abundance is very small: a few parts in $10^4$.  This figure should only be interpreted as an estimate of the magnitude of the 
correction (see text).}
\end{figure}

We have re-run the level code with Eqs.~(\ref{eq:interp1}) and (\ref{eq:interp2}) incorporated and ``turned on'' from $z=1500$ to $3400$.  The change 
in $x_e$ is shown in Fig.~\ref{fig:diff-fl}.  The correction is believed to be most accurate for $z\ge 2200$ when continuum opacity is negligible.  At 
lower redshifts, the corrections of Eqs.~(\ref{eq:interp1}) and (\ref{eq:interp2}) are not reliable.  For Eq.~(\ref{eq:interp1}) this is not a major 
deficiency because at these redshifts feedback [the only process affected by $\pha(\nu_-,z_-)$] is unimportant.  For Eq.~(\ref{eq:interp2}) there is an 
error introduced, however we expect that the change in $x_e$ at $z<2200$ (when continuum opacity is significant) is small because it is only in the far 
damping wings that the corrections described in this section are significant, and continuum opacity makes the line center more important relative to 
the damping wings.  (This is because continuum opacity allows photons to be removed from the line center, whereas without continuum opacity photons can 
only escape the line by redshifting out of the red damping wing.)

The modification to the recombination history resulting from these changes is shown in Fig.~\ref{fig:diff-fl}.  We see that the total effect reaches a 
maximum of $0.03\%$ in the free electron fraction.  This is much smaller than the other effects and comparable to other errors in the code, so we have 
made no attempt to correct for the deviation from Voigt profile or change in $e^{-h\nu/\kB\Tr}$ across the line in the rest of this series of papers.  Since the 
correction is of order the numerical accuracy of the code and involved such a major change to the treatment of the all-important $2^1P^o$--$1^1S$ 
resonance line, we do not claim that the details of Fig.~\ref{fig:diff-fl} are robust; rather we view the results only as confirmation that the effects 
considered are small.

\section{Discussion}
\label{sec:conc}

This was the second paper in a series devoted to cosmological helium recombination.  Here, we examined the problem of two-photon decays in \HeI, 
extending the standard treatment which only accounts for the decay from the $2^1S$ level and ignores the effect of stimulated transitions and 
absorption of the spectral distortion.  We also considered Raman scattering from excited levels in \HeI\ to the ground level ($1^1S$), an effect that 
is distinct from, but closely related to, two-photon decay.  All of these effects change the electron abundance $x_e$ at the level of several 
hundredths of a percent at redshifts $z\ge 1800$.  This results in a change of similar magnitude in the $C_l$s (the precise relation will be quantified 
in more detail in Paper III), which is negligible for cosmic \HeI\ recombination studies.

Our findings regarding the significance of two-photon decays from the $n\ge 3$ levels of \HeI\ differ from some recent statements in the 
literature, most notably Dubrovich \& Grachev \cite{2005AstL...31..359D}, who found a much larger effect.  The main reason for the difference is that 
we find smaller two-photon rates $dA_{2\gamma}/d\nu$ because of destructive interference among different intermediate states in the two-photon 
amplitude (Eq.~\ref{eq:m}) in most parts of the two-photon continuum.  An exception occurs in cases where the two photons emitted in a two-photon decay 
are near an allowed ``1+1'' sequence of decays such as $3^1S\rightarrow 2^1P^o\rightarrow 1^1S$.  These 1+1 decays correspond to resonances in the 
two-photon decay rate at the frequencies corresponding to the one-photon lines (in our example, the $3^1S\rightarrow 2^1P^o$ and $2^1P^o\rightarrow 
1^1S$ lines).  This results in the total (frequency-integrated) rate being very large.  This does not lead to a rapid speed-up of recombination 
however, because the photons emitted in the optically thick resonance lines have a very high probability of re-absorption.  In order to complete the 
recombination calculation it is necessary to split the photons into resonant and nonresonant regions.  The nonresonant regions are handled in the usual 
way for two-photon decays, i.e. they lead to an additional rate that is included in the rate equations.  The two-photon decays in the resonant regions 
are treated as sequences of one-photon decays, with the two-photon effects leading to a modified line profile since with multiple intermediate states 
the Lorentz curve no longer accurately describes the line profile (in the atom's rest frame).  It is essential in this analysis that the treatment of 
the resonant region takes into account the fact that the line is optically thick, otherwise unrealistically fast recombination would be obtained.

The analysis presented in this paper was aimed primarily at helium recombination, however most of the underlying physics is the same for hydrogen 
recombination.  There are two-photon decays from the $n\ge 3$ levels in \HI, and their rates $dA_{2\gamma}/d\nu$ scale as $n^{-3}$ 
\cite{1987PhRvA..36.2155F} for the same reasons described here.  These rates also possess resonances at the frequencies corresponding to 1+1 decays 
such as $3s\rightarrow 2p\rightarrow 1s$.  In general hydrogen recombination matters more for the CMB power spectrum than helium recombination, and in 
particular Wong \& Scott \cite{2006astro.ph.10691W} have found changes in the $C_l$s of several tenths of a percent using rates much smaller than those of DG05.  A 
full calculation for hydrogen would use the two-photon spectra $dA_{2\gamma}/d\nu$, which could be computed by the same methods used here, and take 
into account the modification of the Lyman line profiles due to two-photon corrections.  Such a calculation is beyond the scope of this series of 
papers, but should be a high priority for the CMB community.

\begin{acknowledgments}

C.H. is a John Bahcall Fellow in Astrophysics at the Institute for Advanced Study.
E.S. acknowledges the support from grants NASA LTSAA03-000-0090 and NSF PHY-0355328.  

We acknowledge useful conversations with Jens Chluba, Bruce Draine, Jim Peebles, Doug Scott, Uro\v s Seljak, and Rashid Sunyaev.  
We also thank Joanna Dunkley for critical readings and comments prior to publication.

\end{acknowledgments}

\appendix

\section{Dipole matrix elements for large $n$}
\label{app:largen}

This Appendix evaluates the dipole matrix elements of the form $\langle n^1L||{\bf d}||{n'}^1P^o\rangle$ for large $n$ and small $s=n'-n$ using the Wentzel-Kramers-Brillouin (WKB) method.  This approach is useful since the dipole matrix elements of this form are dominated by large radii where the WKB method works (it breaks down at radii of order $a_0$ or less).  Our goal is to demonstrate the near-exact cancellation of contributions to ${\cal M}_{2\gamma}$ in Eq.~(\ref{eq:m}) that we mentioned in Sec.~\ref{ss:largen}.  We note that WKB-type solutions to the Coulomb approximation wave function have been previously used for several other applications \cite{1980JPhB...13L.101F, 1981OptSp..51...13D}.  The formula presented here is actually equivalent to the special case of Ref.~\cite{1981OptSp..51...13D} in which the eccentricity of the orbit goes to 1, however we provide a simplified derivation here in order to show the fastest route to the key result (Eq.~\ref{eq:0}).

For large $n$, the helium atom can be treated by the Coulomb approximation in which the outer electron (of charge $-e$) moves in the Coulomb potential defined by the combination of the inner electron and nucleus (of charge $+e$).  Except at small $r$, its radial wave function $R(r)$ thus satisfies the Schr\"odinger equation $R''(r)=-k^2(r)R(r)$, where
\beq
k^2(r) = \frac{2m_e}{\hbar^2}\left[ E + \frac{e^2}{r} - \frac{\hbar^2L(L+1)}{2m_er^2} \right].
\eeq
For $E<0$, this equation possesses a classically forbidden region $r>\rmax$, where $\rmax$ is the solution to $k^2(\rmax)=0$.  In the classically allowed region, the WKB solution for $R(r)$ is
\beq
R(r) = (-1)^{n-L-1}\frac{N}{\sqrt{k(r)}}\cos \varphi(r),
\eeq
where the $(-1)^{n-L-1}$ factor is chosen by convention to make the wave function positive near the origin for $N>0$ (it has $n-L-1$ radial nodes) and the radial phase is
\beq
\varphi(r) = -\frac\pi4+\int_r^\rmax k(r)\,dr.
\eeq
The normalization constant is taken to be positive, and to enforce the condition $\int |R(r)|^2\,dr=1$.  For small $L$, we have $\rmax\approx 
e^2/(-E)$, the classically allowed region extends down to $r\ll\rmax$, and then (for small $L$)
\beq
k(r) = \frac e\hbar\sqrt{2m_e(r^{-1}-r_{\rm max}^{-1})}.
\eeq
From this we find
\beq
\int_0^\infty |R(r)|^2\,dr \approx \frac{N^2}2\int_0^\rmax \frac{dr}{k(r)}
= \frac{\pi\hbar r_{\rm max}^{3/2}}{4\sqrt{2m_e}\,e}N^2,
\eeq
\label{eq:getnorm}
where we have replaced $\cos^2\varphi(r)$ with $1/2$ since we integrate over many oscillations of the wave function.  This gives
\beq
R(r) = \frac{(-1)^{n-L-1}2}{(\pi\rmax)^{1/2}(\rmax/r-1)^{1/4}}\cos\varphi(r).
\eeq

In order to compute radial matrix elements with these wave functions for small $s$, we need to consider the effect on the wave function of small changes in $k^2(r)$ resulting from changes in $E$ and $L$.  In general there will be a very small change in $k(r)$, and hence a small change in the amplitude of the solution, but if $s$ is of order a few then we may get a significant change in the phase $\varphi(r)$.  Indeed the phase difference can be written as
\beqa
\Delta \varphi(r) \!\!&=&\!\! \int_r^\rmax \Delta k(r)\,dr + k(\rmax)\Delta\rmax
\nonumber \\
&=&\!\! \int_r^\rmax \frac{\Delta[k^2(r)]}{2k(r)}\,dr.
\eeqa
(The second term goes away because $\rmax$ is a zero of $k^2$.)  The change in $k^2(r)$ has a contribution $2m_e\Delta E/\hbar^2$ if we change the energy, and another contribution $-\Delta[L(L+1)]/r^2$ if we change the angular momentum.  Thus we have
\beqa
\Delta\varphi(r) \!\!&=&\!\! \frac{m_e^{1/2}\Delta E}{\sqrt 2\,e\hbar} \int_r^\rmax \frac{dr}{\sqrt{r^{-1}-r_{\rm max}^{-1}}}
\nonumber \\ &&\!\!
- \frac{\hbar\Delta[L(L+1)]}{2\sqrt{2m_e}\,e}\int_r^\rmax \frac{dr}{r^2\sqrt{r^{-1}-r_{\rm max}^{-1}}}.
\;\;\;\;
\eeqa
It is easy to verify that for $\Delta[L(L+1)]$ of order unity (it is 2 for $S$--$P$ transitions and 4 for $P$--$D$ transitions) and $r/a_0$ greater than a few, the second integral produces a phase shift of $\Delta\varphi(r)\ll 1$.  Therefore we drop it.  The function $\Delta\varphi(r)$ can be solved analytically but is most easily expressed through the following parametric form.  Let us define the dimensionless function
\beq
\tau = 2r_{\rm max}^{-3/2}\int_r^\rmax \frac{dr}{\sqrt{r^{-1}-r_{\rm max}^{-1}}}
\eeq
so that $\Delta\varphi(r) = (m_e^{1/2}r_{\rm max}^{3/2}\Delta E/2\sqrt{2}\,e\hbar)\tau$.  Then by the substitution $r=\rmax(1+\cos\eta)/2$ we can derive $\tau=\eta+\sin\eta$, i.e. the relation between $r$ and $\tau$ is a cycloid function.  Note that $\tau=0$ at $r=\rmax$ and $\tau=\pi$ at $r=0$.

The matrix elements between two levels $n^1L$ and ${n'}^1P^o$ depend on the integral
\beq
R_{n,n',L} = \int R^\ast_{n^1L}(r) R_{{n'}^1P^o}(r)r\,dr;
\eeq
noting that $L$ changes by 1 between the initial and final states, and that for small $s=n'-n$ the normalizations of the wave functions are very similar, we may write
\beqa
R_{n,n',L} \!\!&\approx&\!\! \frac{(-1)^{s-1}4}{\pi\rmax} 
\nonumber \\ && \!\!\times
\int_0^\rmax \!\frac{\cos\varphi(r)\cos[\varphi(r)+\Delta\varphi(r)]}{\sqrt{\rmax/r-1}} r\,dr.
\;\;\;\;\;\;\;\;\;
\eeqa
If we note that $\varphi(r)$ is rapidly varying but $\Delta\varphi(r)$ is not, then the product of cosines can be averaged over several cycles to get
\beq
R_{n,n',L} \approx \frac{(-1)^{s-1}2}{\pi\rmax} \int_0^\rmax r\,\frac{\cos\Delta\varphi(r)}{\sqrt{\rmax/r-1}}dr.
\eeq
Changing variables to $\tau$ gives
\beq
R_{n,n',L} \approx \frac{(-1)^{s-1}}{2\pi}\rmax \int_0^\pi (1+\cos\eta)\cos\Delta\varphi\,d\tau.
\eeq
The integrand is even in $\tau$ so we may extend the range of integration down to $-\pi$ and divide by 2.  We may also replace the cosine by a complex exponential since the imaginary part is odd in $\tau$ and hence vanishes.  This gives
\beq
R_{n,n',L} \approx \frac{(-1)^{s-1}}{4\pi}\rmax \int_{-\pi}^\pi (1+\cos\eta)e^{i\Omega\tau}\,d\tau,
\eeq
where $\Omega = m_e^{1/2}r_{\rm max}^{3/2}\Delta E/2\sqrt 2\,e\hbar$.  Now for large $n$, the energies are given by
\beq
E \approx -\frac{e^2}{2a_0(n+\delta_L)^2},
\eeq
where $\delta_L$ is the quantum defect for angular momentum $L$ \cite{1991PhRvA..44.5448D}.  Note that $\delta_L=0$ for the hydrogenic case, but in 
helium there is a nonzero value due to the complicated physics occuring at small $r$ (of order $a_0$).  The quantum defects for \HeI\ singlets are 
$-0.1397$ ($^1S$), $0.0121$ ($^1P^o$), and $-0.0021$ ($^1D$) \cite{1991PhRvA..44.5448D,2006CaJPh..84...83M}.  Therefore we have $\Delta E \approx 
(e^2/a_0n^3)(s+\delta_1-\delta_L)$ and $\rmax\approx 2a_0n^2$, which implies $\Omega \approx s+\delta_1-\delta_L$.  Thus \beq
R_{n,n',L} \approx \frac{(-1)^{s-1}}{2\pi}a_0n^2 \int_{-\pi}^\pi (1+\cos\eta)e^{i(s+\delta_1-\delta_L)\tau}\,d\tau.
\eeq
Thus we see that the radial matrix element is simply the Fourier transform of the cycloid function.  This is consistent with semiclassical intuition since the cycloid is the classical trajectory of a particle in a Coulombic potential with very small angular momentum.  The reduced matrix element required to compute Eq.~(\ref{eq:m}) is obtained by multiplying by the relevant angular factors:
\beq
\langle n^1L ||{\bf d}||{n'}^1P^o\rangle \approx (-1)^{L_>+s-1}L_>^{1/2} ea_0n^2\fcyc(s+\delta_{1L}),
\label{eq:rmatrix}
\eeq
where we have introduced the shorthand $\delta_{1L}\equiv\delta_1-\delta_L$ and
\beq
\fcyc(\Omega) = \frac1{2\pi}\int_{-\pi}^\pi (1+\cos\eta)e^{i\Omega\tau}\,d\tau.
\label{eq:f}
\eeq

The key result -- the cancellation of contributions to ${\cal M}_{2\gamma}$ for large $n$ -- comes from the following identity:
\beqa
\sum_{s=-\infty}^\infty \!\!\!\!&&\!\!\!\! (-1)^s\fcyc(s+\delta)
\nonumber \\
&=& \!\!
\int_{-\pi}^\pi (1+\cos\eta)\left[\sum_{s=-\infty}^\infty (-1)^se^{i(s+\delta)\tau}\right]\frac{d\tau}{2\pi}
\nonumber \\ &=& \!\!
\int_{-\pi}^\pi (1+\cos\eta)e^{i\delta\tau} \sha\left( \frac\tau{2\pi}+\frac12 \right) \frac{d\tau}{2\pi}
=0,\;\;\;\;\;\;\;\;\;\;
\label{eq:0}
\eeqa
since $1+\cos\eta=0$ when $\tau$ is an odd multiple of $\pi$.  (Here $\sha$ is the sampling function.)

\section{Steady-state line with finite linewidth}
\label{app:ss}

In this Appendix we consider a simple analytic model for the steady-state line profile in the vicinity of a resonance, i.e. an approximate solution to Eq.~(\ref{eq:e0}).  This approximation is valid when the half-width of the ``resonant'' part of the spectrum, $\nucut$, is small compared to the frequency diference between neighboring resonances as well as compared to the thermal scale $\kB\Tr/h$.  As an example, for the \HeI\ $2^1P^o$--$1^1S$ line, we have used $\nucut=58\,$THz; the frequency distance to the next allowed resonance ($3^1P^o$--$1^1S$) is 452 THz; and the thermal scale is $110[(1+z)/2000]\,$THz.

It is easily seen that the matrix element ${\cal M}_{2\gamma}$ for the $i\rightarrow 1^1S$ two-photon process possesses a simple pole at each resonance $\nu,\nu'=\Delta E({n'}^1P^o)/h$.  Therefore the two-photon decay rate, which is the square of the matrix element times phase space factors, can be written in a power series
\beq
\frac{d\Gamma_{2\gamma}}{d\nu} = \sum_{\mu=-2}^\infty q_{i,\mu} \Delta\nu^\mu,
\eeq
where $\Delta\nu=\nu-\nu_{1^1S-{n'}^1P^o}$.  The power series cuts off at $\mu=-2$ because the square of a function with a simple pole can have a pole of no higher than the second order.  It is easy to read off from Eqs.~(\ref{eq:rate-2g}) and (\ref{eq:m}) that the leading term for the $i\rightarrow {n'}^1P^o\rightarrow 1^1S$ pole is
\beqa
q_{i,-2} \!\!&=&\!\! \frac{\alpha^6 \nu^3_{1^1S-{n'}^1P^o}\nu^3_{{n'}^1P^o-i}}{108(2L+1)a_0^6{\cal R}^6}[1+\pha(\nu_{{n'}^1P^o-i})]
\nonumber \\ && \!\!\times
\left| \langle 1^1S||{\bf d}||{n'}^1P^o\rangle \langle {n'}^1P^o||{\bf d}||i\rangle \right|^2,
\eeqa
where we have taken $\pha(\nu_{1^1S-{n'}^1P^o})\ll1$ in the Wien tail of the CMB.  Using the conversion from dipole matrix element to Einstein coefficient, and replacing the phase space density with its blackbody value, this can be re-written as
\beq
q_{i,-2} = \frac{A_{i\rightarrow {n'}^1P^o}A_{{n'}^1P^o\rightarrow 1^1S}}{4\pi^2(1-e^{-h\nu_{{n'}^1P^o-i}/\kB\Tr})}.
\eeq
It follows from this and Eq.~(\ref{eq:l2}) that the absorption cross section to level $i$ is
\beq
\sigma_i^{(2\gamma)} = \sum_{\mu=-2}^\infty Q_{i,\mu} \Delta\nu^\mu,
\eeq
where the leading order term is
\beq
Q_{i,-2} = \frac{c^2g_i}{32\pi^3\nu_{1^1S-{n'}^1P^o}^2}
\frac{A_{i\rightarrow {n'}^1P^o}A_{{n'}^1P^o\rightarrow 1^1S}}{e^{h\nu_{{n'}^1P^o-i}/\kB\Tr}-1}.
\eeq
This could alternatively be written as
\beq
Q_{i,-2} = \frac{c^2g_i}{32\pi^3\nu_{1^1S-{n'}^1P^o}^2}
A_{{n'}^1P^o\rightarrow 1^1S}
\Gamma_{{n'}^1P^o\rightarrow i}.
\label{eq:q-2}
\eeq
A similar argument shows that Eq.~(\ref{eq:q-2}) applies to the resonance in the Raman scattering cross section corresponding to $1^1S\rightarrow {n'}^1P^o\rightarrow i$ as well.

Equation~(\ref{eq:e0}) thus becomes
\beq
\frac{\partial\pha_0}{\partial\nu} = \kappa\left[ \pha_0 - \phaL e^{-h\Delta\nu/\kB\Tr}\right],
\label{eq:e0p}
\eeq
where $\phaL = x_{{n'}^1P^o}/3x_{1^1S}$ and
\beq
\kappa = \frac{\nH c x_{1^1S}\sigma^{(2\gamma+\rm Raman)}}{H\nu}.
\eeq
Expanding $\kappa$ as a power series, $\kappa = \sum_{\mu=-2}^\infty \kappa_\mu \Delta\nu^\mu$, we find that the lowest-order term is
\beqa
\kappa_{-2} \!\!&=&\!\! \sum_i \frac{\nH c x_{1^1S}}{H\nu_{1^1S-{n'}^1P^o}}Q_{i,-2}
\nonumber \\
\!\!&=&\!\! \frac{\nH c x_{1^1S}A_{{n'}^1P^o\rightarrow 1^1S}}{32\pi^3H\nu_{1^1S-{n'}^1P^o}^3}
\sum_i \Gamma_{{n'}^1P^o\rightarrow i}.
\eeqa
In the final expression, the prefactor outside the sum is easily recognized as $\tauS/4\pi^2$, where $\tauS$ is the Sobolev depth through the line.  The sum is the total width of the ${n'}^1P^o$ level (which is the line width $\Gamma_{\rm line}$ of ${n'}^1P^o$--$1^1S$ since the $1^1S$ level has negligible width) times the fraction of transitions from ${n'}^1P^o$ that go to other excited states.  Therefore the sum is $\Gamma_{\rm line}f_{\rm inc}$ and we may write
\beq \kappa_{-2} = \frac{\tauS\Gamma_{\rm line}f_{\rm inc}}{4\pi^2} = \Delta\nu_{\rm line}.
\label{eqn:kappatwo}
\eeq
Note that the coefficient $\kappa$ goes to infinity on resonance.  In principle this should be cut off by the Lorentzian width of the line (i.e. the pole displacement in ${\cal M}_{2\gamma}$), and the resonance will also be widened by the Doppler width of the line.  In practice as long as the line center is optically thick this subtelety does not matter: we will have $\pha=\phaL$ at $\Delta\nu=0$.

Our next objective is to solve Eq.~(\ref{eq:e0p}) for small $\Delta\nu$.  Here we take ``small'' to mean that we can work to first order in $h\Delta\nu/\kB\Tr$ and the correction terms $\{\kappa_\mu\}_{\mu=-1}^\infty$.  We may begin by writing the solution,
\beq
\pha_0 = -X\int X^{-1} \kappa\phaL e^{-h\Delta\nu/\kB\Tr}\,d\nu,
\label{eq:pha0}
\eeq
where $X = \exp\int \kappa\,d\nu$.  The constant of integration in $X$ is arbitrary (it trivially cancels out in obtaining $\pha_0$), while that of the integral in Eq.~(\ref{eq:pha0}) is determined by boundary conditions.  We will separately solve for the $\Delta\nu>0$ and $\Delta\nu<0$ regions since $X$ is singular at $\Delta\nu=0$.  The solution for $X$ is
\beq
X = \exp\left(-\frac{\Delta\nu_{\rm line}}{\Delta\nu} + \kappa_{-1}\ln\frac{\Delta\nu}{\nucut}
+ \sum_{\mu=0}^\infty \kappa_\mu\frac{\Delta\nu^{\mu+1}}{\mu}\right),
\eeq
where the choice of denominator in the logarithm is arbitrary (but $\Delta\nu_+$ is convenient).
Substitution into Eq.~(\ref{eq:pha0}) gives
\beqa
\pha_0 \!\!&=&\!\! -\phaL e^{-\Delta\nu_{\rm line}/\Delta\nu}
\left(\frac{\Delta\nu}{\Delta\nu_+}\right)^{\kappa_{-1}}
e^{\sum_{\mu=0}^\infty \kappa_\mu{\Delta\nu}^{\mu+1}/\mu}
\nonumber\\ && \!\!\times\int 
e^{\Delta\nu_{\rm line}/\Delta\nu}
\left(\frac{\Delta\nu}{\nucut}\right)^{-\kappa_{-1}}
e^{-\sum_{\mu=0}^\infty \kappa_\mu{\Delta\nu}^{\mu+1}/\mu}
\nonumber\\ && \!\!\times
\left( \frac{\Delta\nu_{\rm line}}{\Delta\nu^2} + \sum_{\mu=-1}^\infty \kappa_\mu\Delta\nu^\mu \right)
e^{-h\Delta\nu/\kB\Tr}\,d\nu.
\eeqa
Expanding this to first order in $h\Delta\nu/\kB\Tr$ and $\{\kappa_\mu\}_{\mu=-1}^\infty$ gives
\beqa
\pha_0 \!\!&=&\!\! -\phaL e^{-\Delta\nu_{\rm line}/\Delta\nu} \nonumber\\ &&\!\!\times
\left(\! 1 + \kappa_{-1}\ln\frac{\Delta\nu}{\nucut}
+ \sum_{\mu=0}^\infty \kappa_\mu \frac{\Delta\nu^{\mu+1}}\mu\!\right)
\nonumber\\ && \!\!\times\int 
e^{\Delta\nu_{\rm line}/\Delta\nu}
\frac{\Delta\nu_{\rm line}}{\Delta\nu^2} 
\biggl( 1 - \frac{h\Delta\nu}{\kB\Tr}
- \kappa_{-1}\ln\frac{\Delta\nu}{\nucut}
\nonumber\\ &&
- \sum_{\mu=0}^\infty \kappa_\mu \frac{\Delta\nu{^{\mu+1}}}\mu
+ \sum_{\mu=-1}^\infty \frac{\kappa_\mu\Delta\nu{^{\mu+2}}}{\Delta\nu_{\rm line}}
\biggr)\,d\nu.
\eeqa
The integral can be shown by direct differentiation to evaluate to
\beqa
&&\!\!
-e^{\Delta\nu_{\rm line}/\Delta\nu}
\left( 1 - \kappa_{-1}\ln\frac{\Delta\nu}{\nucut}
- \sum_{\mu=0}^\infty \kappa_\mu \frac{\Delta\nu^{\mu+1}}\mu\!\right)
\nonumber\\&&
-\frac{h\Delta\nu_{\rm line}}{kT_r}E_1\left(\frac{-\Delta\nu_{\rm line}}{\Delta\nu}\right) +C,
\eeqa
where $C$ is a constant of integration and $E_1$ is the exponential integral function.  Therefore, to first order in $\{\kappa_\mu\}_{\mu=-1}^\infty$ and 
$h\Delta\nu/\kB\Tr$,
\beqa
\pha_0 \!\!&=&\!\! \phaL\Biggl[
1 + \frac{h\Delta\nu_{\rm line}}{\kB\Tr}e^{-\Delta\nu_{\rm line}/\Delta\nu} E_1\left(-\frac{\Delta\nu_{\rm line}}{\Delta\nu}\right)
\nonumber\\ &&\!\!
+C e^{-\Delta\nu_{\rm line}/\Delta\nu}
\nonumber\\ &&\times
\left( 1 + \kappa_{-1}\ln\frac{\Delta\nu}{\nucut}
+ \sum_{\mu=0}^\infty \kappa_\mu \frac{\Delta\nu^{\mu+1}}\mu\!\right)
\Biggr].\;\;\;\;
\label{eq:solution}
\eeqa

The photon phase space density on the red side of the line is easiest to obtain: since the term multiplying $C$ in Eq.~(\ref{eq:solution}) goes to infinity as $\Delta\nu\rightarrow 0^-$, we must have $C=0$.  We thus have
\beq
\pha_0(\nu_-) = \phaL\left[
1 + \frac{h\Delta\nu_{\rm line}}{\kB\Tr}e^{\Delta\nu_{\rm line}/\nucut} E_1\left(\frac{\Delta\nu_{\rm line}}{\nucut}\right)
\right].
\label{eq:steadystate}
\eeq
Using the expansion of the exponential integral for small values of the argument, we find that if $\Delta\nu_{\rm line}\ll\nucut\ll \kB\Tr/h$, then
\beq
\pha_0(\nu_-) \approx \phaL\left(
1 + \frac{h\Delta\nu_{\rm line}}{\kB\Tr}\ln\frac{\nucut}{1.78\Delta\nu_{\rm line}}
\right),
\label{eq:asym}
\eeq
where $1.78=e^\gamma$ is the exponential function of Euler's constant.

\bibliography{he2gamma}
\end{document}